\documentclass[article,usenatbib,onecolumn]{mn2e}
\usepackage{amsmath}
\usepackage{amssymb}
\usepackage{graphicx}
\usepackage{color}
\newcommand{\ms}{m_{\star}}
\newcommand{\mi}{m_{\rm i}}
\newcommand{\mo}{m_{\rm o}}
\newcommand{\lambdai}{\lambda_{\rm i}}
\newcommand{\ai}{a_{\rm i}}
\newcommand{\ei}{e_{\rm i}}
\newcommand{\Ii}{I_{\rm i}}
\newcommand{\varpii}{\varpi_{\rm i}}
\newcommand{\Omegai}{\Omega_{\rm i}}
\newcommand{\si}{s_{\rm i}}
\newcommand{\ci}{c_{\rm i}}
\newcommand{\lambdao}{\lambda_{\rm o}}
\newcommand{\ao}{a_{\rm o}}
\newcommand{\eo}{e_{\rm o}}
\newcommand{\Io}{I_{\rm o}}
\newcommand{\varpio}{\varpi_{\rm o}}
\newcommand{\Omegao}{\Omega_{\rm o}}
\newcommand{\aio}{a_{\rm i,o}}
\newcommand{\so}{s_{\rm o}}
\newcommand{\co}{c_{\rm o}}
\newcommand{\MJ}{{\rm M}_{\rm J}}

\newcommand{\Ri}{\left< \mathcal{R}_{\rm i} \right> }
\newcommand{\Ro}{\left< \mathcal{R}_{\rm o} \right> }
\newcommand{\Rio}{\left< \mathcal{R}_{\beta} \right> }
\newcommand{\RDres}{\left< \mathcal{R}_{\rm D}^{\rm res} \right>}
\newcommand{\RDsec}{\left< \mathcal{R}_{\rm D}^{\rm sec} \right>}

\newcommand{\RE}{\left< \mathcal{R}_{\rm E} \right>}
\newcommand{\RI}{\left< \mathcal{R}_{\rm I} \right>}

\newcommand{\DD}{\text{D}}

\newcommand{\ta}{\tau_{a}}
\newcommand{\teo}{\tau_{eo}}
\newcommand{\tei}{\tau_{ei}}
\newcommand{\te}{\tau_{e}}
\newcommand{\sa}{\sqrt{\alpha}}
\newcommand{\dt}[1]{\frac{{\rm d} #1}{{\rm d}\tau}}
\newcommand{\dtt}[1]{\frac{{\rm d} #1}{{\rm d}t}}
\newcommand{\dtts}[1]{\frac{{\rm d}^2 #1}{{\rm d}t^2}}

\begin{document}

\title[Migrating planets in 2:1 mean motion resonance]{Evolution of eccentricity and orbital inclination of migrating planets in 2:1 mean motion resonance}

\author[J. Teyssandier and C. Terquem] {Jean
  Teyssandier$^{1}$\thanks{E-mail: teyssand@iap.fr} and Caroline
  Terquem$^{2,1}$\thanks{E-mail:  caroline.terquem@astro.ox.ac.uk} \\
  $^{1}$ Institut d'Astrophysique de Paris, UPMC Univ Paris 06, CNRS,
  UMR7095, 98 bis bd Arago, F-75014, Paris, France \\
  $^{2}$ Department of Astrophysics, University of Oxford, Keble Road,
  Oxford OX1 3RH, UK}



\maketitle

\label{firstpage}

%
%

\begin{abstract}
  We determine, analytically and numerically, the conditions needed
  for a system of two migrating planets trapped in a 2:1 mean motion
  resonance to enter an inclination--type resonance.  We provide an
  expression for the asymptotic equilibrium value that the
  eccentricity $\ei$ of the inner planet reaches under the combined
  effects of migration and eccentricity damping.  We also show that,
  for a ratio $q$ of inner to outer masses below unity, $\ei$ has to
  pass through a value $e_{\rm i,res}$ of order 0.3 for the system to
  enter an inclination--type resonance.  Numerically, we confirm that
  such a resonance may also be excited at another, larger, value
  $e_{\rm i, res} \simeq 0.6$, as found by previous authors.  A
  necessary condition for onset of an inclination--type resonance is
  that the asymptotic equilibrium value of $\ei$ is larger than
  $e_{\rm i,res}$.  We find that, for $q \le 1$, the system cannot
  enter an inclination--type resonance if the ratio of eccentricity to
  semimajor axis damping timescales $t_e/t_a$ is smaller than 0.2 .
  This result still holds if only the eccentricity of the outer planet
  is damped and $q \lesssim 1$.  As the disc/planet interaction is
  characterized by $t_e/t_a \sim 10^{-2}$, we conclude that excitation
  of inclination through the type of resonance described here is very
  unlikely to happen in a system of two planets migrating in a disc.
\end{abstract}

\begin{keywords}
celestial mechanics -- planetary systems -- planetary systems:
formation -- planetary systems: protoplanetary discs -- planets and
satellites: general
\end{keywords}

%
%

\section{Introduction}

At the time of writing, 98 extrasolar multiple planet systems have
been detected by radial velocity surveys and 420 by the {\em Kepler}
mission (Rowe et al. 2014, Lissauer et al. 2014).  A significant
fraction of these systems contain planet pairs in or near a 2:1 mean
motion resonance (Lissauer et al. 2011, Fabrycky et al. 2012, see
also Petrovich et al. 2013).

Capture in mean motion resonance is thought to be the result of
convergent migration of planets (Snellgrove et al. 2001).  Several
studies (Lee \& Peale 2002, Beaug\'e, Ferraz--Mello \& Michtchenko
2003, Lee 2004, Kley et al. 2005), focussing on the dynamics of the two
planets in 2:1 mean motion resonance in the system GJ~876, were
published soon after the discovery of this system (Marcy et al. 2001).
They assume coplanar orbits and focus on explaining the unusual fact
that the orbits of the two planets librate about apsidal alignment in
this system (as opposed to anti--alignment in the Io--Europa system).
The evolution of planetary systems in mean motion resonance has also
been studied with three--dimensional simulations (Thommes \& Lissauer
2003, Libert \& Tsiganis 2009, Lee \& Thommes 2009).  It has been
found that a system of planets in an eccentricity--type resonance may
enter an inclination--type resonance if the eccentricity of the inner
planet becomes large enough.  In this context, very high orbital
inclinations can be reached starting from nearly coplanar
configurations.

Studies of resonant inclination excitation are important as this
mechanism has been proposed to explain the fact that some extrasolar
planets have an orbit which is inclined with respect to the stellar
equatorial plane.

Other processes which have been put forward to produce such an
inclination include interactions between the planet and a companion
(Fabrycky \& Tremaine 2007, Wu, Murray \& Ramsahai 2007, Chatterjee et
al.  2008, Naoz et al. 2011, Wu \& Lithwick 2011), misalignment of the
disc in which the planet forms (Bate, Lodato \& Pringle 2010, Batygin
2012, Terquem 2013), tilting of the stellar spin axis due to
interaction with the disc (Foucart \& Lai 2011, Lai, Foucart \& Lin
2011) or dynamical relaxation of a population of planets (Papaloizou
\& Terquem 2001).

The studies of inclination--type resonances published so far, which
are numerical, have mapped to some extent the parameter space and have
shown that the onset of resonant inclination excitation depends
sensitively on the eccentricity damping timescale and the ratio of the
planets' masses.

In this paper, we derive analytically a necessary condition for the
onset of inclination--type resonance.  The analysis has to be done to
second order in eccentricities and inclinations.  We also perform
numerical simulations to investigate the regime of high
eccentricities.  We now review eccentricity-- and inclination--type
resonances before giving an outline of the plan of the paper.

\subsection{Eccentricity-- and inclination-- type resonances}

\label{sec:resonances}

In this paper, we are interested in the case where convergent
migration of two planets has led to capture into a 2:1 mean motion
resonance.  First order eccentricity--type resonance involves only the
resonant arguments $2 \lambdao - \lambdai - \varpi_{\rm i,o}$, where
$\lambda_{\rm i,o}$ and $\varpi_{\rm i,o}$ are the mean longitudes and
longitudes of pericenters of the inner and outer planets,
respectively, and is not associated with a variation of the
inclinations (see section~\ref{sec:lpe} below).
Inclination--type resonance is of second order and involves the
resonant arguments $4 \lambdao - 2 \lambdai - 2 \Omega_{\rm i,o}$ and
$4 \lambdao - 2 \lambdai - \Omegai - \Omegao$, where $\Omega_{\rm
  i,o}$ are the longitudes of ascending nodes of the planets.  As
shown by Thommes \& Lissauer (2003), it requires the eccentricity of
the inner planet to reach relatively large values to be excited.

Depending on the eccentricities and masses of the planets, stable
eccentricity--type resonances can be (Beaug\'e, Ferraz--Mello \&
Michtchenko 2003, Lee 2004): \\ (i) symmetric with both resonant
arguments librating about 0$^{\circ}$ (and $\varpio - \varpii$
librating about 0$^{\circ}$, i.e. the apsidal lines are
aligned and conjunction occurs when the planets are near pericenter), \\
(ii) anti--symmetric with the resonant arguments librating about
0$^{\circ}$ and 180$^{\circ}$, respectively (and $\varpio - \varpii$
librating about 180$^{\circ}$, i.e. the apsidal lines are
anti--aligned and conjunction occurs when one planet is near
pericenter and the other near apocenter), \\ (iii) asymmetric with the
resonant arguments librating about angles far from 0$^{\circ}$ and
180$^{\circ}$.

Resonances are stable if the distance between the planets at
conjunctions stays large enough. To understand the physics of
resonance, let us consider the case of conjunction occurring when the
planets are near an apse, i.e. librating about 0$^{\circ}$ or
180$^{\circ}$.  In that case, the tangential force exerted by one
planet onto the other integrates to zero over an orbit and there is no
exchange of angular momentum.  The next conjunction therefore occurs
at the same longitude.  However, if the outer planet is migrating
inward (over a timescale much longer than the orbital period), the
conjunction occurs slightly away from the stable longitude.  In that
case, there is a net tangential force which results in the planets
exchanging angular momentum in such a way that subsequent conjunctions
will be closer to the stable longitude.  Therefore, commensurability
is maintained and the inner planet is pushed inward (Goldreich 1965,
Peale 1976).  It can also be shown that during the migration process
the radial perturbative force between the planets tends to increase
their eccentricities (Lissauer et al.  1984).

If the eccentricities are small, the net tangential force exerted by
one planet onto the other is small and therefore the transfer of
angular momentum between the two planets is weak.  However, in that
case, the radial force is much more effective in changing the
orientation (causing the regression) of the line of apses and
therefore in maintaining the orientation of the apses with the
conjunction longitude (Peale 1976, Greenberg 1977).  The
eccentricity--type resonance therefore exists even for small
eccentricities.

For inclination--type resonances, conjunctions librate about the
longitude of a node of one planet.  There is however no resonance for
low inclinations.  This is because although the orientation of the
lines of nodes can be more easily varied when the inclination is
small, the normal perturbative force between the planets, which tends
to vary the orientation of the lines of nodes, becomes smaller when
the inclination decreases (Greenberg 1977).

\subsection{Plan of the paper}

In section~\ref{sec:theory}, we develop an analysis, valid to second
order in eccentricities and inclinations, of a system of two planets
embedded in a disc and in 2:1 mean motion resonance.  We give an
expression of the disturbing function in section~\ref{sec:df}, write
Lagrange's planetary equations in section~\ref{sec:lpe} and explain
how migration and eccentricity damping are included in
section~\ref{sec:migration}.  We assume that the semimajor axis and
the eccentricity of the outer planet are damped by interaction with
the disc, and consider both the case where the eccentricity of the
inner planet is damped and the case where it is not.  In
section~\ref{sec:ana}, we derive a necessary condition for the onset
of inclination--type resonance.  We find that, for a ratio $q$ of
inner mass to outer mass below unity, the eccentricity of the inner
planet has to reach $e_{\rm i, res} \sim 0.3$ for an inclination--type
resonance to be excited.  This cannot be achieved for $t_{ei} / t_a
<0.2$, where $t_{ei}$ and $t_a$ are the timescales over which the
eccentricity of the inner planet and the semimajor axis of the outer
planet are damped.  We also find that the onset of the
inclination--type resonance requires the eccentricity of the outer
planet to reach a critial value $e_{\rm o, res}$.  In
section~\ref{sec:num}, we present the results of $N$--body
simulations.  We write the equations of motion which are solved in the
code in section~\ref{sec:eom} and give the initial setting in
section~\ref{sec:init}.  In section~\ref{sec:ill}, we describe three
illustrative cases corresponding to three different values of the
eccentricity damping timescale in the case where the eccentricities of
both planets are damped over the same timescale $t_e$.  We compare
numerical and analytical results in section~\ref{sec:comp},
investigate how the onset of inclination--type resonance depends on
$q$ and $t_e/t_a$ in section~\ref{sec:map} and study the influence of
varying parameters in section~\ref{sec:condchoice}.  In
section~\ref{sec:cavity}, we consider the case where eccentricity
damping affects only the outer planet.  The effect of inclination
damping is discussed in section~\ref{sec:idamp}.  These simulations
confirm the analysis and show that there is another, larger, value of
$e_{\rm i, res} \simeq 0.6$ (which was found by Thommes \& Lissauer
2003).  The simulations also show that the onset of inclination--type
resonance requires the eccentricity of the outer planet to reach
$e_{\rm o, res} \sim 0.2$ when $q \lesssim 1 $.  The resonant argument
$\phi_2$ librates about $180^{\circ}$ or $0^{\circ}$ (while $\phi_1$
librates about $0^{\circ}$) depending on whether the system enters an
inclination--type resonance with $e_{\rm i, res} \simeq 0.3$ or 0.6,
respectively.  Finally, in section~\ref{sec:discussion} we summarize and
discuss our results.

%
%

\section{Analysis of the resonance}
\label{sec:theory}

\subsection{Disturbing function}
\label{sec:df}

We consider two planets of masses $\mi$ and $\mo$ orbiting a star of
mass $\ms$.  The subscripts 'i' and 'o' refer to the inner and outer
planets, respectively.  The orbital elements $\lambdai$, $\ai$, $\ei$,
$\Ii$, $\varpii$ and $\Omegai$ denote the mean longitude, semi--major
axis, eccentricity, inclination, longitude of pericenter and longitude
of ascending node of the planet of mass $\mi$, with same quantities
with subscript 'o' for the planet of mass $\mo$.  We suppose that the
two planets are close to or in a 2:1 mean motion commensurability,
i.e. the ratio of the mean motions, $n_{\rm i}/n_{\rm o}$, is close or 
equal to 2.  The dynamics is therefore dominated by the resonant and
secular terms in the disturbing function, since all the other terms are
short--period and average out to zero over the orbital periods.

The perturbing functions for the inner and outer planets can
be written under the form (Murray \& Dermott 1999):

\begin{align}
  \Ri & = \frac{G \mo}{\ao} \left( \RDsec + \RDres +\alpha \RE \right),
  \label{Rpli} \\
  \Ro & =  \frac{G \mi}{\ao} \left( \RDsec + \RDres + \frac{1}{\alpha^2} 
    \RI \right), \label{Rplo}
\end{align}

\noindent where $G$ is the constant of gravitation, $\alpha \equiv \ai /
\ao$, $\RDsec$ and $\RDres$ are the secular and resonant contributions
to the direct part of the disturbing function, respectively, $\RE$ is
the indirect part due to an external perturber and $\RI$ is the
indirect part due to an internal perturber.  The brackets indicate
that the quantities are time--averaged.  Note that there is no secular
contribution to $\RE$ and $\RI$.

When an expansion of the perturbing function in the orbital elements
is carried out, the lowest orders at which eccentricities and
inclinations appear are the first and second, respectively, and at
second order, there are no terms in which eccentricities and
inclinations are coupled.  
To study the inclination--type resonance, second--order terms in 
eccentricities need to be included.  
The expansion of the perturbing function to
second order is (Murray \& Dermott 1999):
\begin{eqnarray}
  \RDsec & = &  K_1 \left( \ei^2 + \eo^2 \right)
  + K_2 \ei \eo \cos 
  \left( \varpii - \varpio \right) +
  K_3 \left( \si^2 + \so^2 \right) + K_4 \si \so \cos 
  \left( \Omegai - \Omegao \right) , \label{RDsec} \\ 
  \RDres & = & \ei f_1 \cos \phi_1 + \eo f_2 \cos \phi_2 + 
  \ei^2 f_3 \cos \phi_3 + \ei \eo f_4 \cos \phi_4 +
  \eo^2 f_5 \cos \phi_5 \nonumber \\
  &  & + \si^2 f_6 
  \cos \phi_6 + \si \so f_7 \cos \phi_7 + \so^2 f_8 \cos \phi_8 , 
  \label{RDres} \\
  \RE & = &  -2 \eo \cos \phi_2 , \label{RE} \\
  \RI & = &  - \frac{1}{2} \eo \cos \phi_2 , \label{RI} 
\end{eqnarray}
\noindent where $s_{\rm i, o}= \sin \left( I_{\rm i, o}/2 \right)$,
and the $f_{i} \; (i=1, \ldots, 8)$ and $K_{i} \; (i=1, \ldots, 4)$ are
expressed in term of the Laplace coefficients and $\alpha$.  Their
expression is given in Appendix~\ref{app:coef}.  The resonant
angles $\phi_i \; (i=1, \ldots, 8)$ are defined by: 
\begin{align}
\phi_1 & =  2\lambdao-\lambdai-\varpii, \label{phi1} \\
\phi_2 & =  2\lambdao-\lambdai-\varpio, \label{phi2} \\
\phi_3 & =  4\lambdao-2\lambdai-2\varpii, \label{phi3} \\
\phi_4 & =  4\lambdao-2\lambdai-\varpio - \varpii, \label{phi4} \\
\phi_5 & =  4\lambdao-2\lambdai-2\varpio, \label{phi5} \\
\phi_6 & =  4\lambdao-2\lambdai-2\Omegai, \label{phi6} \\ 
\phi_7 & =  4\lambdao-2\lambdai-\Omegai-\Omegao, \label{phi7} \\
\phi_8 & =  4\lambdao-2\lambdai-2\Omegao. \label{phi8}
\end{align}

\subsection{Lagrange's planetary equations}
\label{sec:lpe}

When the perturbing function is expanded to second order in the
eccentricities and inclinations, Lagrange equations can be written as
follows:

\begin{align}
  \frac{d a_{\beta}}{dt} & =   \frac{2}{n_{\beta} a_{\beta}}
  \frac{\partial \Rio}{\partial \lambda_{\beta}}, \label{dadt} \\
  \frac{d e_{\beta}}{dt} & =   \frac{-1}{n_{\beta} a_{\beta}^2 e_{\beta}}
  \frac{\partial \Rio}{\partial \varpi_{\beta}}, \\
  \frac{d \varpi_{\beta}}{dt} & =   \frac{1}{n_{\beta} a_{\beta}^2 e_{\beta}}
   \frac{\partial \Rio}{\partial e_{\beta}}, \\
  \frac{d \lambda_{\beta}}{dt} & =   n_{\beta} + 
  \frac{1}{n_{\beta} a_{\beta}^2} \left(
    -2 a_{\beta} \frac{\partial \Rio}{\partial a_{\beta}}
    +\frac{e_{\beta}}{2 }
    \frac{\partial \Rio}{\partial e_{\beta}} 
   + \tan \frac{I_{\beta}}{2} \frac{\partial \Rio}{\partial I_{\beta}} 
\right) , 
  \label{dlambdadt} \\
  \frac{d \Omega_{\beta}}{dt} & =  
  \frac{1}{n_{\beta} a_{\beta}^2  \sin I_{\beta}}
  \frac{\partial \Rio}{\partial I_{\beta}}, \\
  \frac{d I_{\beta}}{dt} & = 
  \frac{-1}{n_{\beta} a_{\beta}^2 \sin I_{\beta}} 
  \frac{\partial \Rio}{\partial \Omega_{\beta}}, \label{dIdt}
\end{align}

\noindent where $\beta={\rm i, o}$.  In equation~(\ref{dlambdadt}),
the derivative with respect to $a_{\beta}$ has to be carried out by
ignoring the fact that the angles $\phi_i \; (i=1, \ldots, 8)$ depend
on $a_{\beta}$ through $\lambdai$ and $\lambdao$ (Roy 1978).  We define a
timescale $T$ such that:

\begin{equation}
T = n_{{\rm o} 0}^{-1} \; \frac{\ms}{\mi},
\label{timeT}
\end{equation}

\noindent where the subscript '0' denotes initial value. We consider
planets with masses $m_{\rm i,o} \ll \ms$, so that the mean motions are
approximated by $n_{\rm i,o}= \left( G \ms /a_{\rm i,o}^3 \right)^{1/2}$.

Using the expression of the perturbing function given by
equations~(\ref{Rpli})--(\ref{RI}), we can rewrite Lagrange
equations~(\ref{dadt})--(\ref{dIdt}) in the following dimensionless
form:

\begin{align}
  \frac{d \ai}{d \tau} & =  {\cal C}_{\rm i} 2 \ai 
  \left[ \ei F_1 + \eo F_2  - \eo 2 \alpha  \sin \phi_2
    + F_3 + {\cal{O}}(3) \right], \label{daidt0} \\
\frac{d \ao}{d \tau} & =  {\cal C}_{\rm o}
(-4 \ao ) \left[ \ei F_1 + \eo F_2 - \frac{\eo}{2 \alpha^2} \sin \phi_2 
+ F_3
+ {\cal{O}}(3) \right],
     \label{daodt0}\\
  \frac{d \ei}{d \tau} & =   {\cal C}_{\rm i}
  \left[ -F_1  + {\cal{O}}(2)  \right], 
\label{deidt0}\\
\frac{d \eo}{d \tau} & =  {\cal C}_{\rm o}
\left[ - F_2  + \frac{1}{2 \alpha^2} \sin \phi_2  + {\cal{O}}(2) \right], 
\label{deodt0}\\
  \frac{d \varpii}{d \tau} & =  {\cal C}_{\rm i} \left[ \frac{1}{\ei} \;
  f_1 \cos\phi_1  +2 f_3 \cos \phi_3 + \frac{\eo}{\ei} \; f_4 \cos \phi_4
  + 2K_1 \right. \nonumber \\ 
  &  \left. \hspace*{8.cm} +K_2 \frac{\eo}{\ei} \; 
  \cos \left( \varpii - \varpio \right) + {\cal{O}}(1)
  \right], \label{dvarpiidt} \\
  \frac{d \varpio}{d \tau} & =  {\cal C}_{\rm o}
  \left[ \frac{1}{\eo} \;
    \left( f_2 - \frac{1}{2 \alpha^2} \right) \cos\phi_2 
   + \frac{\ei}{\eo} f_4 \cos \phi_4
    +2 f_5 \cos \phi_5 + 2 K_1
  \right. \nonumber \\ 
  & \left. \hspace*{8.cm} 
    +K_2 \frac{\ei}{\eo}  \cos \left( \varpii - \varpio \right) 
    + {\cal{O}}(1) \right] , \label{dvarpiodt} \\
 \frac{d \lambdai}{d \tau} & =   {\cal C}_{\rm i}
 \left[ \frac{1}{\alpha} \frac{\ms}{\mo} + \frac{1}{2} \ei f_1 \cos\phi_1
+4 \alpha \eo \cos \phi_2 - \Delta + {\cal{O}}(2) \right] ,\\
   \frac{d \lambdao}{d \tau} & =  {\cal C}_{\rm o}
    \left[ \frac{\ms}{\mi} 
     + 2 \ei f_1 \cos\phi_1 +
     \frac{1}{2} \eo \left( 5 f_2 + \frac{3}{2 \alpha^2} \right)  \cos \phi_2 
+ \Delta + {\cal{O}}(2) \right],  \\
   \frac{d \Omegai}{d \tau} & =   \frac{{\cal C}_{\rm i}}{2} 
   \left[ K_3 + \frac{1}{2} \frac{\so}{\si} 
     \left( K_4 \cos( \Omegai - \Omegao ) + f_7 \cos \phi_7 \right)+
     f_6 \cos \phi_6 + {\cal{O}}(1) \right], \label{dOmegaidt} \\
   \frac{d \Omegao}{d \tau} & =   
    \frac{{\cal C}_{\rm o}}{2}
   \left[ K_3  +
     \frac{1}{2} \frac{\si}{\so} \left( K_4 \cos( \Omegai - \Omegao )  
       + f_7 \cos \phi_7 \right) + f_8 \cos \phi_8  + {\cal{O}}(1) \right],
   \label{dOmegaodt}\\
   \frac{d \Ii}{d \tau} & =  {\cal C}_{\rm i}
   \left[ \frac{1}{2} \frac{\so}{\ci} \left( K_4 \sin( \Omegai - \Omegao ) 
       - f_7 \sin \phi_7 \right) - \frac{\si}{\ci} f_6 \sin \phi_6 
   + {\cal{O}}(2) \right] ,
   \\
   \frac{d \Io}{d \tau} & =   
   {\cal C}_{\rm o}
   \left[ - \frac{1}{2} \frac{\si}{\co} \left( K_4 \sin( \Omegai - \Omegao ) 
       + f_7 \sin \phi_7 \right) - \frac{\so}{\co} f_8 \sin \phi_8 
   + {\cal{O}}(2) \right],
   \label{dIodt}
\end{align}

\noindent where  $\tau = t/T$, $c_{\rm i, o}= \cos \left( I_{\rm i,
    o}/2 \right)$, ${\cal C}_{\rm i}=
  \left( a_{{\rm o} 0}/\ao \right)^{3/2} \mo/(\mi \sqrt{\alpha})$,
${\cal C}_{\rm o}=
  \left( a_{{\rm o} 0}/\ao \right)^{3/2}$,
\begin{align}
  F_1 & =   f_1 \sin \phi_1 + 2 \ei f_3 \sin \phi_3 + \eo f_4 \sin \phi_4
  -K_2 \eo \sin \left( \varpii - \varpio \right) , \\
  F_2 & =  f_2 \sin \phi_2 + 
  2 \eo f_5 \sin \phi_5 + \ei f_4 \sin \phi_4
  +K_2 \ei \sin \left( \varpii - \varpio \right) , \\
  F_3 & =  2 \si^2 f_6 \sin \phi_6 + 2 \si \so f_7 \sin \phi_7
  + 2 \so^2 f_8 \sin \phi_8
\end{align}

\noindent and:

\begin{equation}
\Delta = 2 \alpha \left( 
\ei \frac{ \partial f_1}{\partial \alpha} \cos \phi_1
+ \eo \frac{ \partial f_2}{\partial \alpha} \cos \phi_2 \right).
\end{equation}

Note that, with a perturbing function expanded to second order in
eccentricities and inclinations, the rate of change of $\varpi_{\rm
  i,o}$ and $\Omega_{\rm i,o}$ can be obtained only to zeroth order in
eccentricities and inclinations.

The only orbital elements which are relevant when the orbits are
coplanar are $a_{\rm i,o}$, $e_{\rm i,o}$, $\varpi_{\rm i,o}$ and
$\lambda_{\rm i,o}$.  At the order at which the above equations
are valid, we note that their evolution is decoupled from that of
$\Omega_{\rm i,o}$ and $I_{\rm i,o}$.  
Similarly, when the semimajor axes do not vary
significantly, the variations of $\Omega_{\rm i,o}$ and $I_{\rm i,o}$
are decoupled from that of the other orbital elements.

\subsection{Modelling of migration and eccentricity damping}
\label{sec:migration}

We are interested in the case where the two planets capture each
other in a mean motion resonance, which happens for instance if the
outer planet migrates inward due to its interaction with the disc at
least as fast as the inner planet.  To model this, we artificially
decrease the semimajor axis of the outer planet by adding a damping
term $-\ao/\ta$ on the right--hand side of equation~(\ref{daodt0}),
where $\ta$ is the (dimensionless) migration timescale that can be
freely specified.  In principle, due to its interaction with the disc,
the inner planet also migrates independently of the outer one.  For
simplicity, this is not taken into account here.  It would not change
the results presented in this paper as long as the inner planet were
not migrating faster than the outer one.  In this context, the
timescale $\ta$ can be thought of as the timescale over which the outer
planet migrates with respect to the inner one.  Note that a
diminution of $\ai$, although not being added artificially, will be
induced by that of $\ao$ (see below).

Interaction with the disc also leads to damping of the planets
eccentricities.  This is taken into account by adding a damping term
$-\ei /\tei$ and $-\eo / \teo$ on the right--hand side of equations
(\ref{deidt0}) and (\ref{deodt0}), respectively, where $\tei$ and
$\teo$ are the (dimensionless) eccentricity damping timescales.
Eccentricity damping does in turn contribute to the damping of the
semimajor axis by a term $-2 \aio e_{\rm i,o}^2 / \tau_{\rm ei,o}$
(see appendix~\ref{sec:forces}).

With migration and eccentricity damping taken into account,
equations~(\ref{daidt0})--(\ref{deodt0}) become:

\begin{align}
  \frac{d \ai}{d \tau} & =  {\cal C}_{\rm i} 2 \ai 
  \left[ \ei F_1 + \eo F_2  - \eo 2 \alpha  \sin \phi_2
    + F_3 + {\cal{O}}(3) \right] - \frac{2 \ai \ei^2}{\tei} , 
  \label{daidt} \\
\frac{d \ao}{d \tau} & =  {\cal C}_{\rm o}
(-4 \ao ) \left[ \ei F_1 + \eo F_2 - \frac{\eo}{2 \alpha^2} \sin \phi_2 
+ F_3
+ {\cal{O}}(3) \right] - \frac{\ao}{\ta} - \frac{2 \ao \eo^2}{\teo} ,
     \label{daodt}\\
  \frac{d \ei}{d \tau} & =   {\cal C}_{\rm i}
  \left[ -F_1  + {\cal{O}}(2)  \right] - \frac{\ei}{\tei} , 
\label{deidt}\\
\frac{d \eo}{d \tau} & =  {\cal C}_{\rm o}
\left[ - F_2  + \frac{1}{2 \alpha^2} \sin \phi_2  + {\cal{O}}(2) \right]
- \frac{\eo}{\teo} . 
\label{deodt}
\end{align}

\noindent In this paper, we will consider either the case $\tei=\teo$
or the case of infinite $\tei$ (no damping on the inner planet) and
finite $\teo$. Inclination damping has not been included, as it is not relevant here (see
section~\ref{sec:idamp}).

\subsection{Necessary condition for the onset of inclination--type resonance}
\label{sec:ana}

In this section, we derive a necessary condition for a pair of planets
in 2:1 mean motion resonance to enter an inclination--type resonance.
We assume that the planets have captured each other in an exact
resonance, so that $\alpha \equiv \ai/\ao = 2^{-2/3}$ remains constant
in time.  In that case, it is useful to note that $2\alpha =
1/(2\alpha ^2)$.  We also define the mass ratio $q \equiv \mi/\mo$.

In this context, 
equations~(\ref{daidt}) and (\ref{daodt}) can be combined to give:

\begin{equation}
\frac{2}{\ai} \frac{ d \ai}{d \tau}  + \frac{1}{q \sa} \frac{1}{\ao} 
\frac{ d \ao}{d \tau} = - \frac{1}{ q \sa \ta} -  
 4 \frac{\ei^2}{\tei} - \frac{2}{q \sa} \frac{\eo^2}{\teo} .
\end{equation}

\noindent Using $\alpha=\ai/\ao$, we obtain the following
differential equation for $\ai$ and $\ao$:

\begin{equation}
  \left( 1 + 2 q \sqrt{\alpha}  \right) \frac{1}{\aio} 
  \frac{d \aio}{d \tau} = 
  -  \frac{1}{\tau_a} - 
   4 q \sa \frac{\ei^2}{\tei} -2 \frac{\eo^2}{\teo}  .
\label{eq:daiodt}
\end{equation}

\noindent To first order in eccentricities, the solutions are:

\begin{equation}
\aio (\tau)  =   a_{{\rm i} 0, {\rm o} 0} \;
e^{-\tau/ \left[ (1+2q\sa) \ta \right]} .
\label{eq:at} 
\end{equation}

\noindent This shows that, when $q=0$, i.e., when the inner planet is
treated as a test particle, both planets migrate at a rate $\ta$. When
$q$ is nonzero, the migration is slightly slowed down by the mutual
interaction between the two planets.

\subsubsection{Eccentricity--type resonance}
\label{sec:eccres}

As a first step, we calculate the equilibrium values reached by the
eccentricities under the combined effects of migration and
eccentricity damping when the system is in an eccentricity--type
resonance with zero inclinations.

Equations~(\ref{deidt}) and~(\ref{deodt}) can be combined to give:

\begin{equation}
\ei \frac{d \ei}{d \tau} + \frac{\ei^2}{\tei} + 
\frac{1}{q \sa} \left( \eo \frac{d \eo}{d \tau} + \frac{\eo^2}{\teo} \right)
= - {\cal C}_{\rm i} \left[ \ei F_1 + \eo F_2 
- \frac{\eo}{2 \alpha^2} \sin \phi_2 +  {\cal O}(3)\right]
\label{deiodt1}
\end{equation}

\noindent When the inclinations are zero, $F_3 = 0$.  Using
equations~(\ref{daodt}) and~(\ref{eq:daiodt}), we
can then rewrite equation~(\ref{deiodt1}) under the form:

\begin{equation}
\dt{\ei^2} + \frac{ 1+q \sa }{1 + 2 q \sa} \frac{ 4 \ei^2}{\tei} + 
\frac{1}{q\sa} \left[ \dt{\eo^2}+\frac{ 1 + q \sa }{1+2q \sa}
  \frac{ 2 \eo^2}{\teo} \right]  = 
\frac{1}{\ta(1+2q\sa)} .
\label{eq:deieo}
\end{equation}

\noindent When $q \to 0$, $\eo$ has to satisfy $ d \eo^2 / d \tau + 2
\eo^2 / \teo =0$, which means that $\eo$ is damped to zero whereas, in
this limit, $\ei$ reaches the equilibrium value $[\tei/(4
\ta)]^{1/2}$.  

\paragraph{Finite values of $\tei$ and $\teo$ :}

When the eccentricities of both planets are damped and $q \le 1$, i.e.
the outer planet is at least a massive as the inner one, we expect the
eccentricity of the outer planet to grow less than that of the inner
planet.  We therefore neglect the terms involving $\eo$ in
equation~(\ref{eq:deieo}), which then leads to the following
equilibrium value (corresponding to $\tau \gg \tei$) for $\ei$:

\begin{equation}
e_{{\rm i, eq} } = \frac{1}{2} \left( \frac{\tei/\ta}{1+q\sa} \right)^{1/2}.
\label{eq:eiq}
\end{equation}

In figure~\ref{fig:eres}, we plot $e_{{\rm i, eq} }$ as a function of
$q$ for $\tei/\ta=1$, 0.2 and 0.1. The numerical simulations presented in the next section confirm that
neglecting $\eo$ in equation~(\ref{eq:deieo}) is a valid approximation
for the parameters of interest in this paper.  The numerical values
found for $e_{{\rm i, eq} }$ also agree very well with that given by
equation~(\ref{eq:eiq}), as indicated in Figure~\ref{fig:eeqnum}.

\paragraph{Finite value of $\teo$ and $\tei \to \infty$ :}

In some conditions that will be discussed in section~\ref{sec:cavity},
damping due to tidal interaction with the disc acts only on the
eccentricity of the outer planet, not on that of the inner planet.  In
that case, as the planets are in resonance, $\ei$ still reaches an
equilibrium value.  This is shown by equation~(\ref{deidt0}) in which
$F_1=0$ when the resonant angles librate about $0^{\circ}$ or
$180^{\circ}$.  By substituting $d\ei^2/d\tau=0$ and $\tei \to \infty$
in equation~(\ref{eq:deieo}), we obtain the following equilibrium
value (corresponding to $\tau \gg \teo$) for $\eo$:

\begin{equation}
  e_{{\rm o, eq} } =  \left[ \frac{q \sa \; \teo/\ta}{2 \left( 1+q\sa \right)} 
  \right]^{1/2}.
\label{eq:eoq}
\end{equation}

In figure~\ref{fig:eores}, we plot $e_{{\rm o, eq} }$ as a function of
$q$ for $\teo/\ta=0.2$, 0.05 and 0.01. The numerical values found for $e_{{\rm o, eq} }$ agree very well
with that given by equation~(\ref{eq:eoq}), as indicated in
Figure~\ref{fig:notorque}.

\subsubsection{Inclination--type resonance}
\label{sec:incres}

We now calculate the values $e_{\rm i, res}$ and $e_{\rm o, res}$ that
$\ei$ and $\eo$, respectively, have to reach for the inclinations to
start growing. A necessary condition for the system to enter an
inclination--type resonance when both $\tei$ and $\teo$ are finite is
$e_{\rm i, res}<e_{\rm i, eq}$, where $e_{\rm i, eq}$ is given by
equation~(\ref{eq:eiq}).  When $\tei \to \infty$, a necessary condition 
is $e_{\rm o, res}<e_{\rm o, eq}$, where $e_{\rm o, eq}$ is given by
equation~(\ref{eq:eoq}).

Inclination--type resonance can be reached only after an
eccentricity--type resonance has been established (Thommes \& Lissauer
2003) and therefore all the resonant angles librate about some values.
Due to eccentricity damping, libration of the resonant angles is
actually offset from a fixed value (generally 0$^{\circ}$ or
$180^{\circ}$) by a term on the order of $e_{\rm i}/\tei$ or $e_{\rm
  o}/\teo$ (Goldreich \& Schlichting 2014). Because of the $1/\tei$ or
$1/\teo$ factor, this term is much smaller than those retained in the
analysis here, so that we will ignore it.  Therefore the
time--derivative of the resonant angles is close to zero.
Equations~(\ref{phi1}), (\ref{phi2}), (\ref{phi6}) and (\ref{phi8})
then yield:

\begin{equation}
\dt{\varpii} = \dt{\varpio}= \dt{\Omegai} = \dt{\Omegao},
\end{equation}

\noindent i.e., the nodes and pericenters all precess at the same
rate.  In the inclination--type resonance we consider here, as shown
in the numerical simulations presented below, the resonant angles
$\phi_6$, $\phi_7$ and $\phi_8$ librate about $180^{\circ}$,
$0^{\circ}$ and $180^{\circ}$, respectively, and $\Omegai-\Omegao$
librates about $180^{\circ}$, in agreement with Thommes \& Lissauer
(2003).  In the analysis that follows, we therefore use
$\phi_7=0^{\circ} $, $\phi_6=180^{\circ}$, $\phi_8 =180^{\circ}$
and $\Omegai-\Omegao=180^{\circ}$.

We first write $d \Omegai/ d \tau = d \Omegao/ d \tau$ by equating
equations~(\ref{dOmegaidt}) and~(\ref{dOmegaodt}).  This leads to an
algebraic equation for $\si/\so$.  Using the fact that $f_6=f_8$ and
$K_3-f_8=(-K_4+f_7)/2$ (see appendix~\ref{app:coef}), we can write the
solution of this equation as:

\begin{equation}
\label{eq:siso}
\frac{\so}{\si} = q \sa.
\end{equation}

\noindent For values of $q$ less than unity, the inclination of the
inner planet is more easily excited than that of the outer planet.
Note that the ratio of the inclinations does not depend on $\ta$ nor
$\tei$ or $\teo$.

We now write $d \varpio/ d \tau = d \Omegao/ d \tau$ by equating
equations~(\ref{dvarpiodt}) and~(\ref{dOmegaodt}) and $d \varpii/ d
\tau = d \Omegai/ d \tau$ by equating equations~(\ref{dvarpiidt})
and~(\ref{dOmegaidt}).  In the resulting set of equations, we replace
$\si / \so$ by the value found above.  This set of two equations can
now be solved to calculate $e_{\rm i, res}$ and $e_{\rm o, res}$,
which are the values of $\ei$ and $\eo$, respectively, when the
inclination--type resonance is reached.  This calculation requires the
knowledge of $\phi_1$ and $\phi_2$ at the time when the
inclination--type resonance is triggered.  For two planets in 2:1 mean
motion resonance, it has been found (e.g. Lee \& Peale 2002) that, in
general, $(\phi_1, \phi_2)=(0^{\circ},180^{\circ})$ or
$(0^{\circ},0^{\circ})$, depending on whether the eccentricities are
small or large, respectively ('large' meaning that second order terms
in the eccentricities are not negligible).  Fixing $\phi_1=0^{\circ}$
and $\cos^2 \phi_2=1$, we obtain:

\begin{eqnarray}
  e_{\rm i, res} & = & \frac{-f_1 \Delta_3 + \left( f_2 - \frac{1}{2 \alpha^2} 
\right) \left(f_4 +K_2 \right)}
{-\left( f_4 +K_2 \right)^2  + \Delta_3 
\Delta_4},
\label{eires_small} \\
e_{\rm o, res} & = & \frac{f_1 \left( f_4 +K_2 \right)- 
\Delta_4 \left( f_2 - \frac{1}{2 \alpha^2} \right)}{\Delta_3 
\Delta_4 - \left( f_4 +K_2 \right)^2} \cos \phi_2 ,
\label{eores_small}
\end{eqnarray}

\noindent where we have defined $\Delta_1=K_3+q \sqrt{\alpha} (f_7
-K_4)/2 -f_6$, $\Delta_2=K_3+(f_7 -K_4)/(2 q \sqrt{\alpha})-f_8$,
$\Delta_3=2 f_5 + 2 K_1 - \Delta_2/2$ and 
$\Delta_4=2 f_3 + 2 K_1 - \Delta_1/2$.  

We note that $e_{\rm i, res}$ and $e_{\rm o, res}$ do not depend on
the damping timescales.  When $q \rightarrow 0$,
equations~(\ref{eires_small}) and~(\ref{eores_small}) give $e_{\rm i,
  res} = f_1/\left[ (K_3-f_6)/2 -2(f_3+K_1) \right] \simeq 0.22$ and
$e_{\rm o, res}=0$, respectively.

Equation~(\ref{eores_small}) gives a positive value of the
eccentricity for $\phi_2=0^{\circ}$ but a negative value for
$\phi_2=180^{\circ}$.  When $\phi_2$ evolves from libration about
$0^{\circ}$ to libration about $180^{\circ}$, the apsidal lines evolve
from being aligned to being anti--aligned, i.e. the pericentre of the
outer orbit is changed to apocentre.  In that case, for the equation
of the outer ellipse which was used in deriving Lagrange's equations
to be preserved, $\eo$ should be changed to $-\eo$.  The value of
$\eo$ corresponding to $\phi_2=180^{\circ}$ is
therefore the opposite as that given by equation~(\ref{eores_small}).

\begin{figure}
\begin{center}
\includegraphics[scale=0.7]{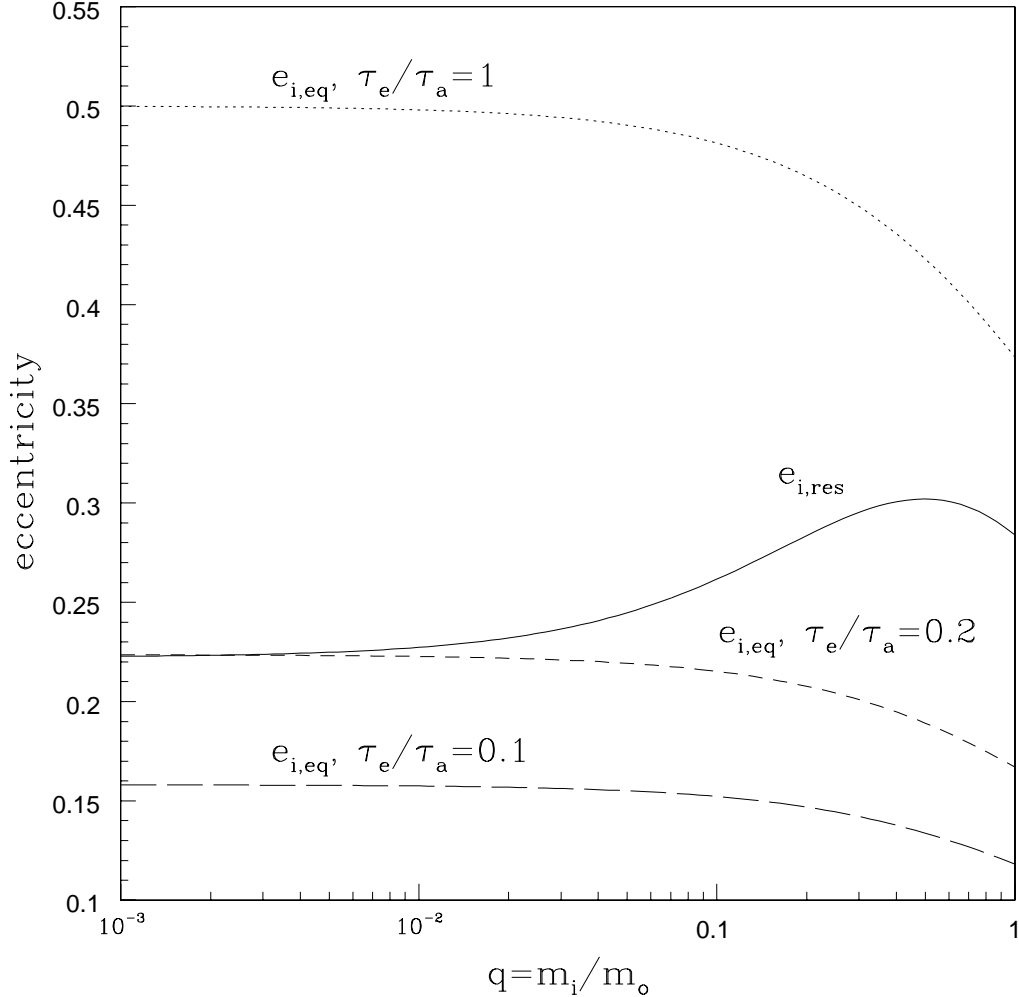}
\end{center}
\caption{{\em Solid line:} Eccentricity $e_{\rm i, res}$ of the inner
  planet when the system enters an inclination--type resonance {\em
    versus} $q$ in logarithmic scale.  This curve corresponds to
  equation~(\ref{eires_small}) with $\cos^2 \phi_2=1$. Also shown is
  the equilibrium eccentricity for $\tei/\ta=0.1$ ({\em long dashed line}),
  $\tei/\ta=0.2$ ({\em short dashed line})
  and $\tei/\ta=1$ ({\em dotted line}).  These curves correspond to
  equation~(\ref{eq:eiq}).  The system cannot enter an inclination--type 
  resonance when $\tei/\ta <0.2$.}
\label{fig:eres}
\end{figure}

\begin{figure}
\begin{center}
\includegraphics[scale=0.7]{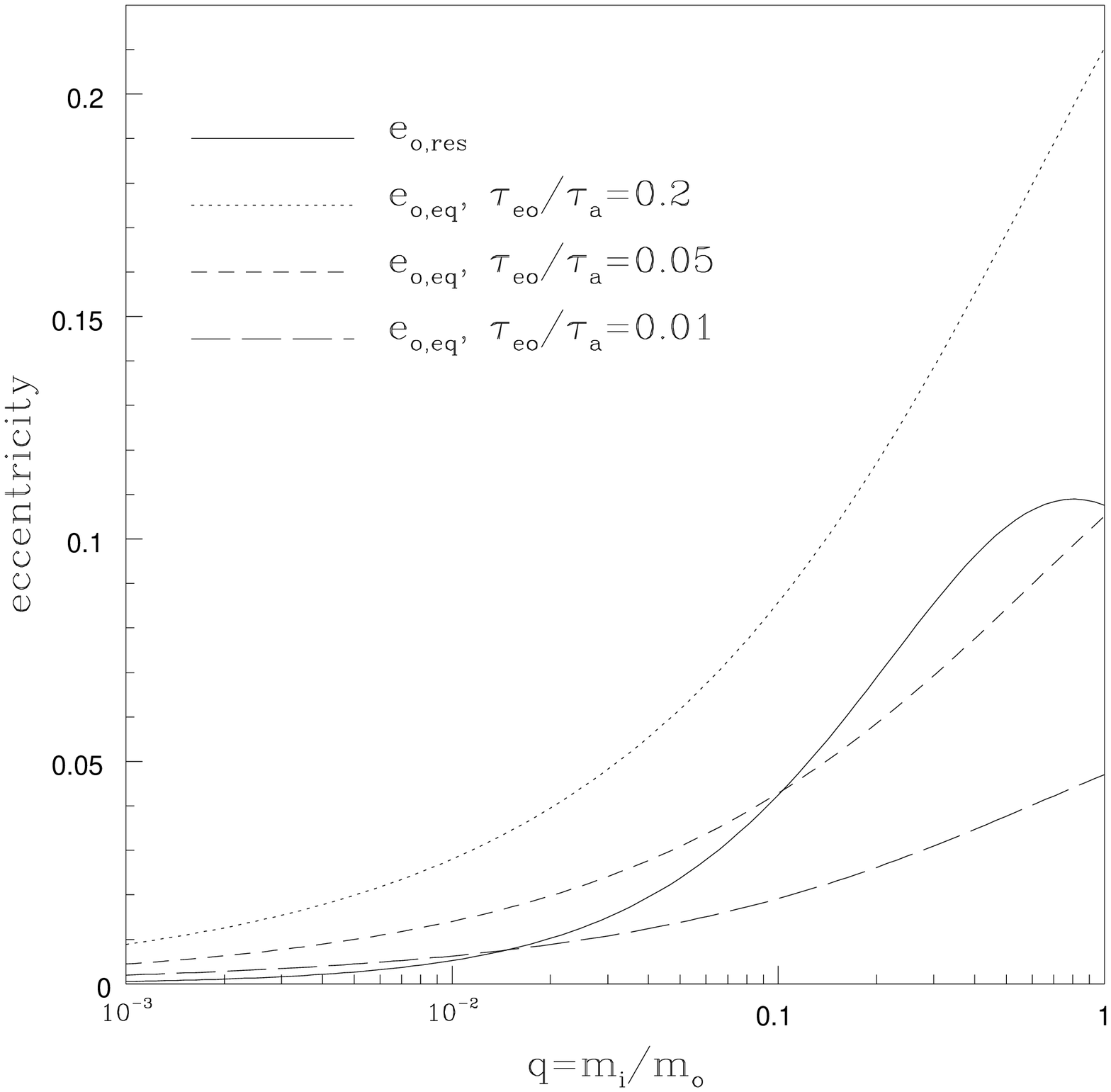}
\end{center}
\caption{{\em Solid line:} Eccentricity $e_{\rm o, res}$ of the outer
  planet when the system enters an inclination--type resonance {\em
    versus} $q$ in logarithmic scale.  This curve corresponds to
  equation~(\ref{eores_small}) with $\cos^2 \phi_2=1$. Also shown is
  the equilibrium eccentricity for $\tei \to \infty$ and
  $\teo/\ta=0.01$ ({\em long dashed line}), $\teo/\ta=0.05$ ({\em
    short dashed line}) and $\teo/\ta=0.2$ ({\em dotted line}).  These
  curves correspond to equation~(\ref{eq:eoq}).  Note that the value
  of $e_{\rm o, res}$ found numerically for $q \lesssim 1$ is almost
  twice as large as that displayed here, being $\simeq 0.2$. In this
  context, and for $q \lesssim 1$, the system cannot enter an
  inclination--type resonance when $\teo/\ta < 0.2$. }
\label{fig:eores}
\end{figure}

In figure~\ref{fig:eres}, we plot the eccentricity $e_{\rm i, res}$
given by equation~(\ref{eires_small}) as a function of $q$.  For $q$
varying between 0.1 and 1, we see that $e_{\rm i, res} \simeq 0.3$.
The eccentricity $e_{\rm o, res}$ given by equation (\ref{eores_small})
is shown in figure~\ref{fig:eores}.  We see that $e_{\rm o, res}$
varies much more with $q$ than $e_{\rm i, res}$.

The numerical results presented in the next section show that there
are two regimes in which the system may enter an inclination--type
resonance: (i) a {\em small eccentricity} regime, which is described
by the analysis above (although the value of $e_{\rm o,res}$ found
numerically is about twice as large as that found analytically), and
(ii) a {\em large eccentricity} regime, in which inclination--type
resonance is obtained for $e_{\rm i, res} \simeq 0.6$ and $e_{\rm o,
  res} \simeq 0.2$ for $q \lesssim 1$.  The large eccentricity regime
cannot be captured by the analysis above, as our equations are not
developed up to a sufficient order in $e_{\rm i,o}$.  As noted above,
by expanding the perturbing function to second order in $e_{\rm i,o}$,
we obtain $d \varpi_{\rm i,o} / dt$ and $d \Omega_{\rm i,o} / dt$ to
zeroth order only in $e_{\rm i,o}$.

When both $\tei$ and $\teo$ are finite, we can find the values of $q$
and $\tei/\ta$ for which inclination--type resonance may occur by
writing the necessary condition $e_{\rm i, res}< e_{\rm i, eq} $.
Using equations~(\ref{eq:eiq}) and~(\ref{eires_small}) and ignoring
the (weak) variation of $e_{\rm i, res}$ with $q$, this leads to:

\begin{equation}
\label{eq:condres}
 \frac{\tei/ \ta}{1+q\sa} > 4 e_{\rm i, res}^2.
\end{equation}

\noindent Note that the above condition is necessary but not
sufficient, as we have not taken into account the fact that $\eo$ must
also reach $e_{\rm o, res}$ (we cannot write this condition as we have
not calculated $e_{\rm o,eq}$ when $\tei$ is finite).
Figure~\ref{fig:eres} shows that the system cannot enter an
inclination--type resonance when $\tei/\ta < 0.2$.  When $q$ decreases,
the inclination--type resonance is reached for shorter eccentricity
damping timescales.  This is expected as eccentricity is pumped up to
higher values when $q$ is smaller. We will discuss condition~(\ref{eq:condres}) in more details in the next
section, after deriving $e_{\rm i, res}$ from the numerical
simulations.

When $\tei \to \infty$, a necessary condition for the system to enter
an inclination--type resonance is $e_{\rm o, res}<e_{\rm o, eq}$.  As
can be seen from figure~\ref{fig:eores}, and if we adopt for $q
\lesssim 1$ the value of $e_{\rm o, res} \simeq 0.2$ found
numerically, which is about twice as large as the value derived
analytically, we find that the system cannot enter an
inclination--type resonance when $\teo/\ta < 0.2$.

%
%

\section{Numerical simulations}
\label{sec:num}

The analysis presented above is valid only for small eccentricities.
We now perform numerical simulations of a pair of migrating planets in
the vicinity of a 2:1 mean motion resonance to extend the study to
arbitrarily high eccentricities.

\subsection{Equations of motion}
\label{sec:eom}

We consider a system consisting of a primary star and two planets
initially embedded in a gaseous disc surrounding the star. The planets
undergo gravitational interaction with each other and the star and are
acted on by tidal torques from the disc.  To study the evolution of
the system, we use the $N$--body code described in Papaloizou~\&
Terquem~(2001) in which we have added the effect of the disc torques
(see also Terquem \& Papaloizou 2007).

The equations of motion for the inner and outer planets are:
\begin{align}
\dtts{\textbf{r}_{\rm i}}  & = 
-\frac{G\ms\textbf{r}_{\rm i}}{\lvert \textbf{r}_{\rm i}\rvert^3} 
- \frac{Gm_{\rm o}(\textbf{r}_{\rm i}-\textbf{r}_{\rm o})}
{\lvert \textbf{r}_{\rm i}-\textbf{r}_{\rm o} \rvert ^3} - 
\sum_{\gamma={\rm i,o}}\frac{Gm_{\gamma}\textbf{r}_\gamma}
{\lvert \textbf{r}_\gamma \rvert ^3} + 
\mbox{\boldmath$\Gamma$}_{\rm d,i} + 
\mbox{\boldmath$\Gamma$}_{\rm GR,i}, 
\label{eq:eomi} \\
\dtts{\textbf{r}_{\rm o}}  & =  
-\frac{G\ms\textbf{r}_{\rm o}}{\lvert \textbf{r}_{\rm o}\rvert^3} 
- \frac{Gm_{\rm i}(\textbf{r}_{\rm o}-\textbf{r}_{\rm i})}
{\lvert \textbf{r}_{\rm o}-\textbf{r}_{\rm i} \rvert ^3} - 
\sum_{\gamma={\rm i,o}}\frac{Gm_{\gamma}\textbf{r}_\gamma}
{\lvert \textbf{r}_\gamma \rvert ^3} + 
\mbox{\boldmath$\Gamma$}_{\rm d,o} + 
\mbox{\boldmath$\Gamma$}_{\rm GR,o}, \label{eq:eomo}
\end{align}

\noindent where $\textbf{r}_{\rm i}$ and $\textbf{r}_{\rm o}$ denote
the position vector of the inner and outer planets, respectively, and
we have included the acceleration of the coordinate system based on
the central star (third term on the right--hand side).  Acceleration
due to tidal interaction with the disc is dealt with through the
addition of extra forces as in Papaloizou \& Larwood (2000) (see also
Terquem \& Papaloizou 2007):
\begin{align}
\mbox{\boldmath$\Gamma$}_{\rm d,i} & =  
- \frac{2}{t_{ei} \lvert\textbf{r}_{\rm i} \rvert^2}
\left( \dtt{\textbf{r}_{\rm i}} \cdot \textbf{r}_{\rm i} \right)
\textbf{r}_{\rm i}, 
\label{gammadi} \\
\mbox{\boldmath$\Gamma$}_{\rm d,o} & =   
- \frac{2}{t_{eo} \lvert\textbf{r}_{\rm o} \rvert^2}
\left( \dtt{\textbf{r}_{\rm o}} \cdot \textbf{r}_{\rm o} \right)
\textbf{r}_{\rm o}
-\frac{1}{t_{m}}\dtt{\textbf{r}_{\rm o}}, 
\label{gammado}
\end{align}

\noindent where $t_{m}$, $t_{eo}$ and $t_{ei}$ are the timescales over
which the angular momentum of the outer planet, its eccentricity and
the eccentricity of the inner planet, respectively, are damped through
tidal interaction with the disc.  As mentioned in
section~\ref{sec:migration}, the $1/t_{m}$ term is applied to the
outer planet only.  We show in appendix~\ref{sec:forces} that these
forces do indeed lead to exponential damping of the eccentricities and
outer semimajor axis.  Note however that the migration timescale $t_m$
defined through equation (\ref{gammado}) is equal to twice the
semimajor axis damping timescale defined through
equation~(\ref{daodt}).  In other words, $t_{m}=2 t_{a}$, where $t_{a}
= \tau_a T$ and $T$ is given by equation~(\ref{timeT}).  For the
eccentricity damping timescales, we simply have that $t_{ei}$ and
$t_{eo}$ defined through equations~(\ref{gammadi}) and~(\ref{gammado})
are equal to $\tei T$ and $\teo T$, respectively, where $\tei$ and
$\teo$ are the dimensionless damping timescales defined in
section~\ref{sec:migration}.

Relativistic effects are included through the following acceleration:
\begin{equation}
  \mbox{\boldmath$\Gamma$}_{\rm GR,i,o} =
  -\frac{6G^2\ms^2}{c^2}\frac{\textbf{r}_{\rm i,o}}{\vert 
    \textbf{r}_{\rm i,o} \vert^4},
\end{equation}
with $c$ being the speed of light. It is found that this term, 
which induces a precession of the arguments of pericenters, 
does not affect significantly the results. This is because the 
planets do not approach the star closely enough. 
For this reason also, tidal interaction between the planets and the star 
can be neglected.
 
The effect of inclination damping due to planet--disc interaction,
which also has not been included here, will be discussed in
section~\ref{sec:idamp} below.

\subsection{Initial setting}
\label{sec:init}

In all the runs presented below, unless mentioned otherwise, we
consider a one Jupiter mass outer planet ($m_{\rm o} = 1$~$\MJ$) and
start the two planets slightly outside a 2:1 MMR at $ a_{{\rm i} 0}
=3.05$~AU and $a_{{\rm o} 0} = 5$ AU, corresponding to a period ratio
of 2.1.  The initial eccentricities and inclinations are small:
$e_{{\rm i0}}=e_{{\rm o0}}=5 \times 10^{-3}$, $I_{{\rm i0}}=
0.01^{\circ}$ and $I_{{\rm o0}}=0.02^{\circ}$. The longitudes of
pericenters and of ascending nodes are all equal to $0^{\circ}$
initially.  We fix $t_m=1.4 \times 10^{6}$~years which corresponds to
a semimajor axis damping timescale $t_a=7 \times 10^{5}$~years.  This
is roughly equal to the disc evolution timescale at $3$~AU, as should
be the case for type II migration.  To start with, we fix for
simplicity $t_{ei}=t_{eo} \equiv t_e$, i.e. $\tei=\teo \equiv \te$,
and consider different values of $t_e$ ranging from $\sim 10^5$ to
$\sim 10^6$~years.  In section~\ref{sec:cavity}, we will consider
$t_{ei} \to \infty$ and finite values of $t_{eo}$. The dependence of
our results on the choice of initial conditions will be discussed in
section \ref{sec:condchoice}.


\subsection{Illustrative cases}
\label{sec:ill}

In this section, we illustrate the main features of the dynamical
evolution of two planets migrating in 2:1 mean motion resonance as a
function of the damping eccentricity timescale.  We consider strong,
moderate and weak eccentricity damping, corresponding to $t_e/t_a
\equiv \tau_e/\tau_a=0.25$, 0.8 and 4, respectively.  Here the inner
planet has a mass $\mi =0.7$~$\MJ$, so that the mass ratio is $q=0.7$.

\subsubsection{Strong eccentricity damping: No inclination--type resonance}

\begin{figure}
\begin{center}
\includegraphics[scale=0.7]{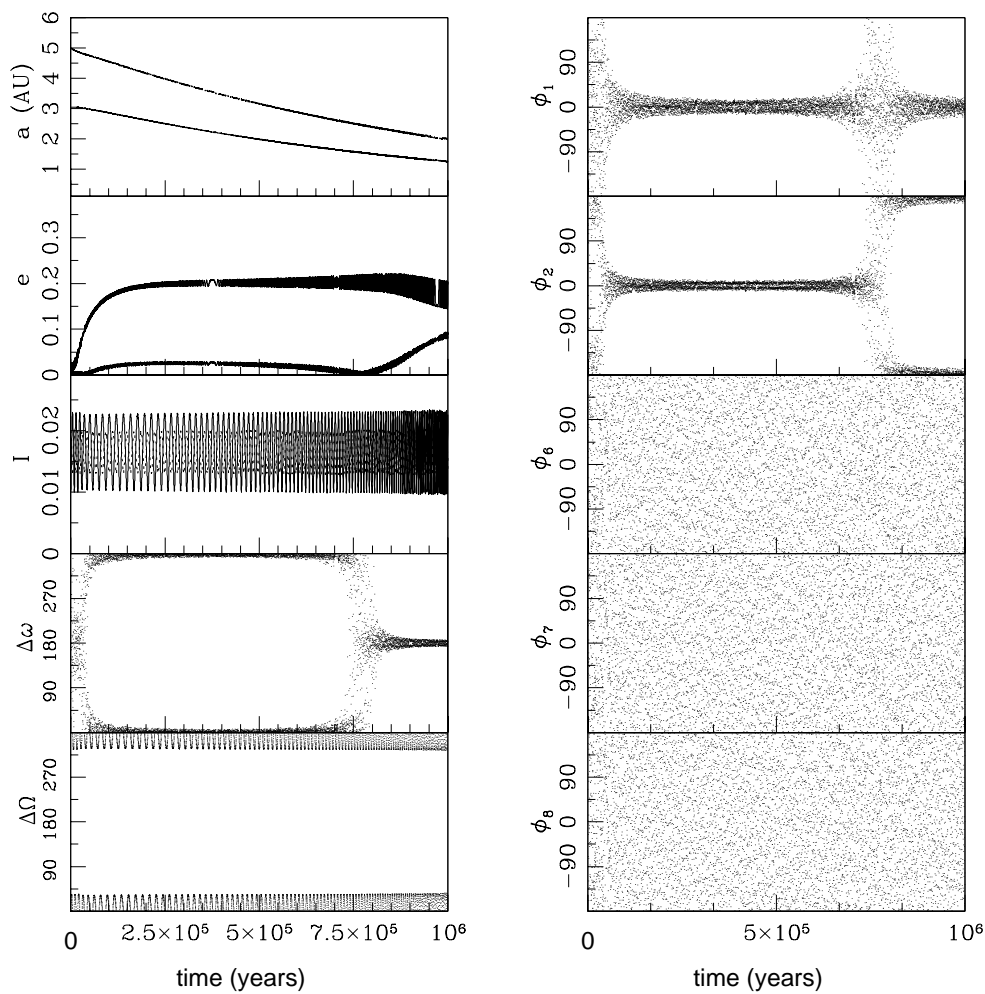}
\end{center}
\caption{Evolution of a system in 2:1 mean motion resonance for
  $q=0.7$ and $\te/\ta=0.25$ in the case $\tei=\teo \equiv \te$.
  \textit{Left column, from top to bottom:} Semi--major axes (in AU),
  eccentricities and inclinations of the two planets, $\Delta \varpi =
  \varpii-\varpio$ and $\Delta \Omega =\Omegai-\Omegao$ {\it versus}
  time (in years). \textit{Right column, from top to bottom:} resonant
  angles $\phi_1$, $\phi_2$, $\phi_6$, $\phi_7$ and $\phi_8$ {\it
    versus} time (in years).  All the angles are given in degrees.  In
  the plot displaying the eccentricities, the upper and lower curves
  represent $\ei$ and $\eo$, respectively.  Shortly after the
  beginning of the simulation, the planets are captured into an
  eccentricity--type resonance: $\ei$ grows until it reaches $e_{\rm
    i, eq} = 0.2$ and $\Delta \varpi$, $\phi_1$ and $\phi_2$ librate
  about 0$^{\circ}$.  After $t \simeq 8 \times 10^5$~years, $\ei$
  starts decreasing while $\eo$ gets larger, and the value about which
  $\phi_2$ librates switches rather abruptly to $180^{\circ}$.
  Throughout the evolution, $\ei$ stays too small to allow for
  inclination--type resonance.  Accordingly, the inclinations remain
  small, the resonant angles $\phi_6$, $\phi_7$ and $\phi_8$ behave
  chaotically and $\Delta \Omega=0$ throughout the evolution. }
\label{fig:strongte}
\end{figure}

We first consider the case $t_e=1.75 \times 10^{5}$~years,
corresponding to $\te/\ta =0.25$.  Figure~\ref{fig:strongte} shows the
evolution of $a_{\rm i,o}$, $e_{\rm i,o}$, $I_{\rm i,o}$, $\Delta
\varpi = \varpii - \varpio$, $\Delta \Omega = \Omegai - \Omegao$,
$\phi_1$, $\phi_2$, $\phi_6$, $\phi_7$ and $\phi_8$ between 0 and
$10^6$~years.  Very quickly after the beginning of the simulation, the
planets capture each other in a 2:1 eccentricity--type resonance.
From equation~(\ref{eq:eiq}), we expect the eccentricity of the inner
planet to reach an equilibrium value $e_{\rm i, eq} = 0.2$, in good
agreement with the value observed in the simulation.  At first, the
resonant angles $\phi_1$ and $\phi_2$ librate about $0^{\circ}$.
However, after about $8 \times 10^5$~years, as the planets get closer
to each other during their convergent migration, $\eo$ starts to
increase significantly while $\ei$ decreases, and the value about
which $\phi_2$ librates switches rather abruptly to $180^{\circ}$,
which is indicative of a regime with smaller eccentricities.  As
expected from the analysis, $\ei$ does not become large enough for an
inclination--type resonance to develop.  Accordingly, the inclinations
remain small and the resonant angles $\phi_6$, $\phi_7$ and $\phi_8$,
associated with inclination--type resonances, behave chaotically.

\subsubsection{Moderate eccentricity damping: Inclination--type
  resonance with small eccentricities}

\begin{figure}
\begin{center}
\includegraphics[scale=0.7]{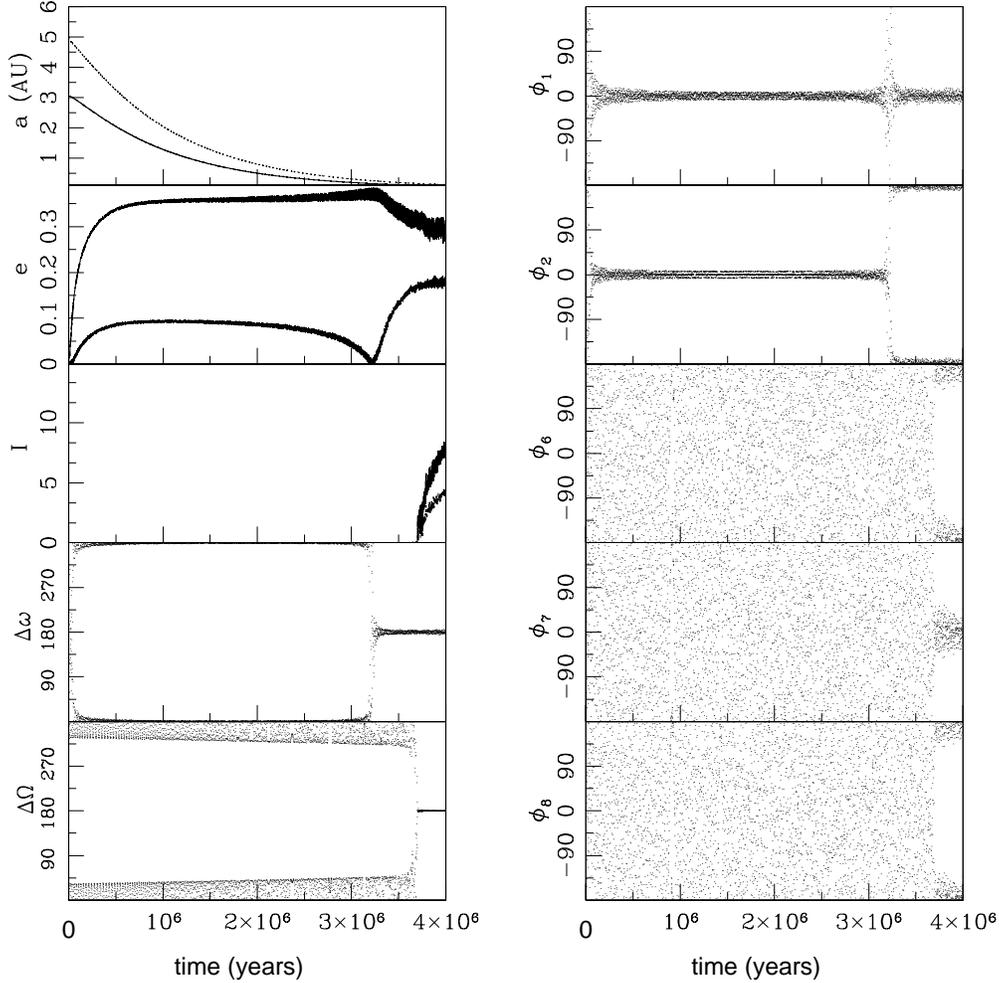}
\end{center}
\caption{Same as figure~\ref{fig:strongte} but for $\te/\ta=0.8$.  In
  the plots displaying the eccentricities and inclinations, the upper
  curves represent $\ei$ and $\Ii$, respectively, whereas the lower
  curves represent $\eo$ and $\Io$, respectively.  Here again, shortly
  after the beginning of the simulation, the planets are captured into
  an eccentricity--type resonance: $\ei$ grows until it reaches
  $e_{\rm i, eq} = 0.35$ and $\Delta \varpi$, $\phi_1$ and $\phi_2$
  librate about 0$^{\circ}$.  After $t \simeq 3.2 \times 10^6$~years,
  $\ei$ starts decreasing while $\eo$ gets larger, and the value about
  which $\phi_2$ librates switches to $180^{\circ}$.  When $\ei \simeq
  0.3$, an inclination--type resonance starts to develop.
  Accordingly, the resonant angles $\phi_6$, $\phi_7$ and $\phi_8$
  start librating about $180^{\circ}$, $0^{\circ}$ and $180^{\circ}$,
  respectively, while $ \Delta \Omega$ librates about $180^{\circ}$.
  The inclinations grow quickly.  }
\label{fig:modte}
\end{figure}

We now consider the case $t_e=5.6 \times 10^{5}$~years, corresponding
to $\te/\ta =0.8$.  The evolution of $a_{\rm i,o}$, $e_{\rm i,o}$,
$I_{\rm i,o}$, $\Delta \varpi$, $\Delta \Omega$ and the resonant
angles between 0 and $4 \times 10^6$~years are displayed in
figure~\ref{fig:modte}.  In this case, equation~(\ref{eq:eiq})
predicts that the eccentricity of the inner planet should reach an
equilibrium value $e_{\rm i, eq} = 0.35$, in good agreement with the
numerical result.  We see that, at first, $\ei$ grows very quickly,
until it reaches the equilibrium value.  At the same time, $\phi_1$
and $\phi_2$ librate about $0^{\circ}$.  Like in the previous case,
after a certain time (here $t \simeq 3.2 \times 10^6$~years), $\ei$
starts decreasing while $\eo$ gets larger, and the value about which
$\phi_2$ librates switches to $180^{\circ}$.  
When $\ei \simeq 0.3$, an inclination--type resonance starts to
develop, in very good agreement with the expectation from
equation~(\ref{eires_small}).  Accordingly, the resonant angles
$\phi_6$, $\phi_7$ and $\phi_8$ start librating about $180^{\circ}$,
$0^{\circ}$ and $180^{\circ}$, respectively, while $ \Delta \Omega$
librates about $180^{\circ}$.  The inclinations grow quickly,
maintaining a ratio in agreement with equation~(\ref{eq:siso}). The
inclination--type resonance is triggered when $\eo \simeq 0.18$, which
is a bit larger than the value of $0.1$ predicted by
equation~(\ref{eores_small}).

Note that, in this particular case, the inclination--type resonance
occurs at $t \sim 3.5 \times 10^{6}$ yr, which is likely to be longer
than the disc's lifetime.

We observe that the system does not enter an inclination--type
resonance when $\ei$ first reaches 0.3, at about $t=2 \times
10^5$~years.  This is because at that point $\eo$ is still smaller
than $e_{\rm o,res}$.  For $q <1$, $\eo$ can reach $e_{\rm o,res}$ only
when the two planets get close enough during their convergent
migration for the perturbation onto the outer planet to become
significant.

\subsubsection{Weak eccentricity damping: Inclination--type resonance
  with large eccentricities}

\begin{figure}
\begin{center}
\includegraphics[scale=0.7]{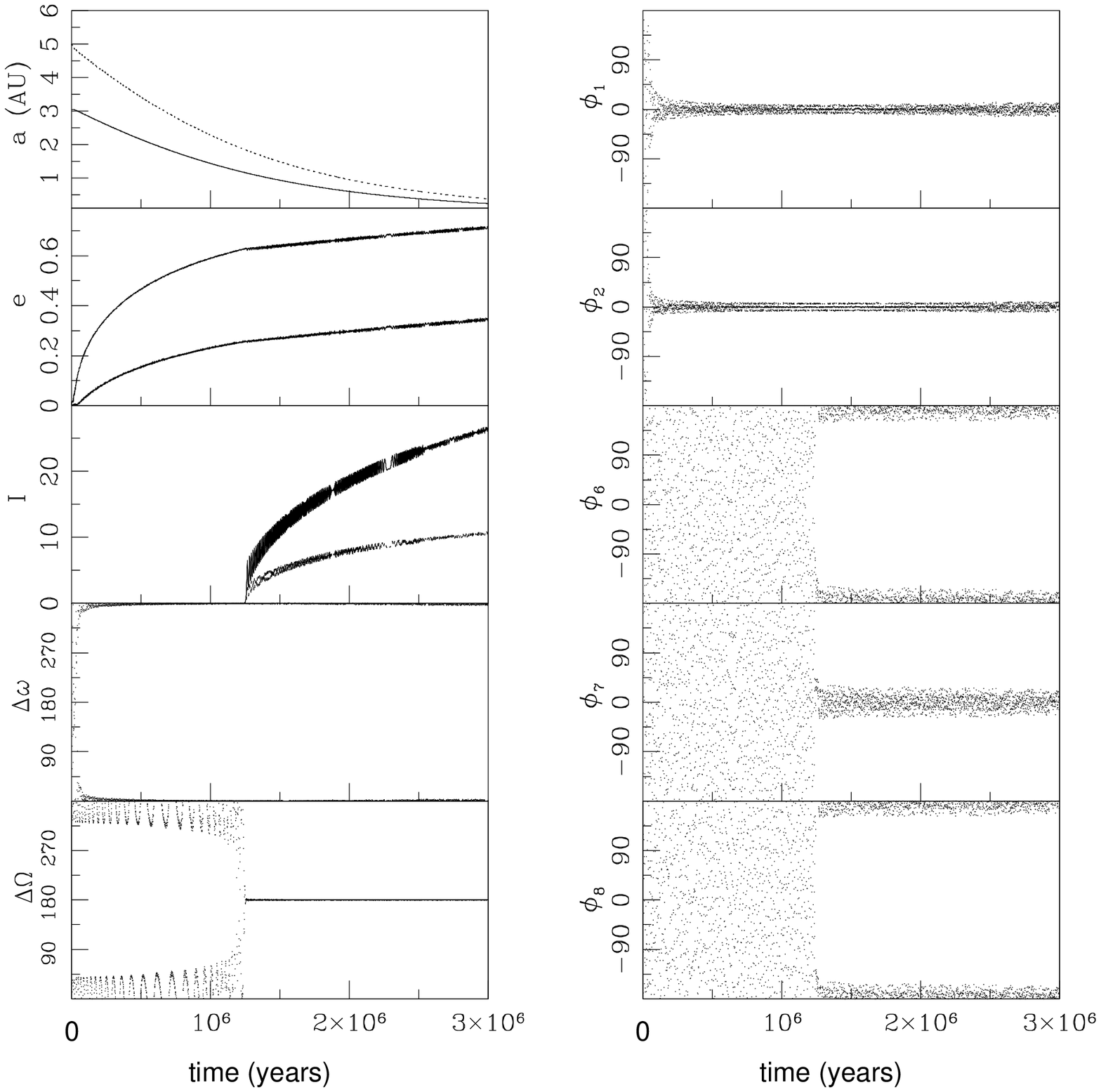}
\end{center}
\caption{Same as figure~\ref{fig:strongte} but for $\te/\ta=4$.  In
  the plots displaying the eccentricities and inclinations, the upper
  curves represent $\ei$ and $\Ii$, respectively, whereas the lower
  curves represent $\eo$ and $\Io$, respectively.  Here again, shortly
  after the beginning of the simulation, the planets are captured into
  an eccentricity--type resonance and $\ei$ and $\eo$ grow.  In the
  present case, $\phi_1$ and $\phi_2$ librate about $0^{\circ}$
  throughout the simulation.  When $\ei \sim 0.6$, the system enters
  an inclination--type resonance and $\Delta \Omega$, $\phi_6$,
  $\phi_7$ and $\phi_8$ start librating about $180^{\circ}$,
  $180^{\circ}$, $0^{\circ}$ and $180^{\circ}$, respectively, while
  the inclinations start growing.  }
\label{fig:weakte}
\end{figure}

We finally consider the case $t_e=2.8 \times 10^{6}$~years,
corresponding to $\te/\ta =4$.  The evolution of $a_{\rm i,o}$,
$e_{\rm i,o}$, $I_{\rm i,o}$, $\Delta \varpi$, $\Delta \Omega$ and the
resonant angles between 0 and $3 \times 10^6$~years are displayed in
figure~\ref{fig:weakte}.  In this case, equation~(\ref{eq:eiq})
predicts that the eccentricity of the inner planet should reach an
equilibrium value $e_{\rm i, eq} = 0.8$.  As this is very large, the
analysis presented in section~\ref{sec:ana} is probably not valid.  
In any case, equilibrium is not attained during the time of the simulation.
Here, $\phi_1$ and $\phi_2$ librate about $0^{\circ}$ throughout the
simulation, which is expected for a regime with high eccentricities.
Both $\ei$ and $\eo$ grow from the beginning of the simulation.  When
$\ei \simeq 0.6$, at $t \simeq 1.2 \times 10^6$~years, the system
enters an inclination--type resonance, characterized by $\Delta
\Omega$, $\phi_6$, $\phi_7$ and $\phi_8$ librating about
$180^{\circ}$, $180^{\circ}$, $0^{\circ}$ and $180^{\circ}$,
respectively.  This case is similar to those investigated by Thommes
\& Lissauer (2003).

We observe that the system does not enter an inclination--type
resonance when $\ei$ reaches 0.3 at about $t=2 \times 10^5$~years, as
$\eo$ is still smaller than $e_{\rm o,res}$ at that time.

In the simulation reported here, the inclination of the outer planet
stays smaller than the disc aspect ratio, so that migration is not
interrupted.  However, we expect the inclination to become larger at
some later time, so that the outer planet loses contact with the disc
and stops migrating.

\subsection{Comparison with analytical results}
\label{sec:comp}

In this section, we compare the numerical values of $e_{\rm i,eq}$,
$e_{\rm i,res}$, $e_{\rm o,res}$ and of the ratio $\so / \si$ when the
system is in an inclination--type resonance, with the analytical
values given by equations~(\ref{eq:eiq}), (\ref{eires_small}),
(\ref{eores_small}) and~(\ref{eq:siso}), respectively.

\subsubsection{Equilibrium eccentricity of the inner planet}
\label{sec:eieq}

In section~\ref{sec:ana}, we derived an analytical expression for the
equilibrium value of $\ei$ assuming $\eo$ to be negligible compared to
$\ei$.  As can be seen from figures~\ref{fig:strongte},
\ref{fig:modte} and~\ref{fig:weakte}, this approximation is at least
marginally valid.

\begin{figure}
\begin{center}
\includegraphics[scale=0.7,trim=0cm 2cm 0cm 3cm, clip=true]{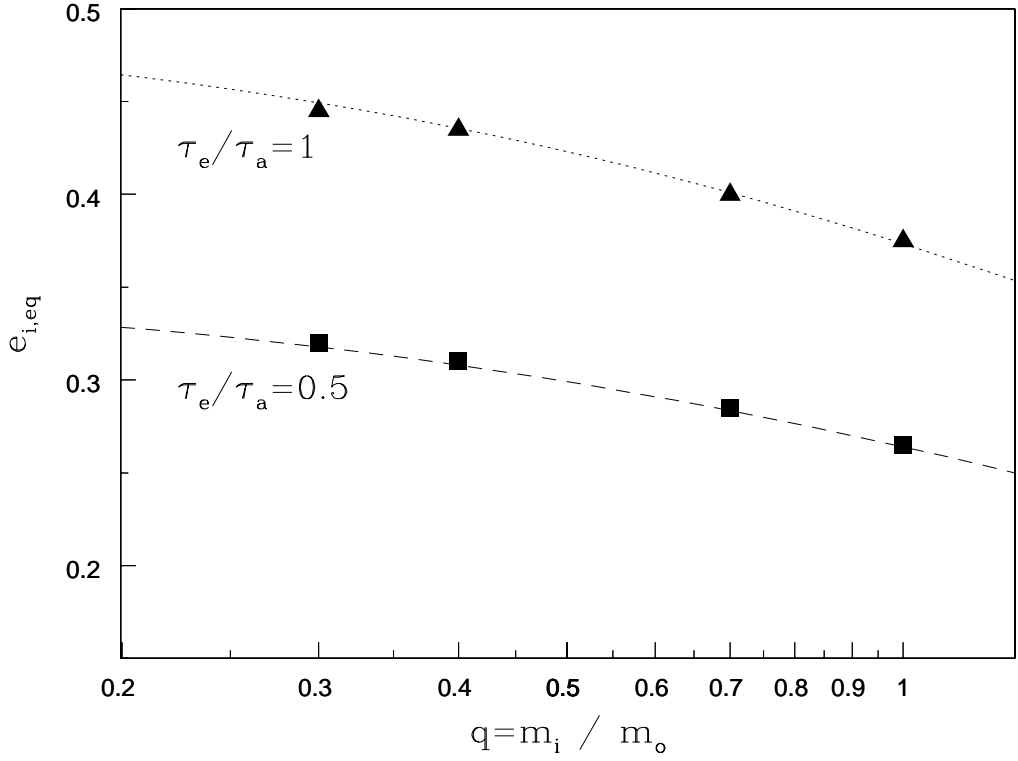}
\end{center}
\caption{Equilibrium eccentricity of the inner planet {\em versus}
  mass ratio $q$ for $\te/\ta=1$ {\em (dotted line, triangles)}, and
  $\te/\ta=0.5$ {\em (dashed line, squares)} in the case $\tei=\teo
  \equiv \te$. The lines represent $e_{\rm i, eq}$, calculated using
  eq.~(\ref{eq:eiq}) whereas the symbols represent the numerical
  values.}
\label{fig:eeqnum}
\end{figure}

In figure~\ref{fig:eeqnum}, we compare the analytical value of $e_{\rm
  i,eq}$ given by equation~(\ref{eq:eiq}) with the results of the
numerical integration.  We consider two eccentricity damping
timescales, corresponding to $\te/\ta=0.5$ and $1$, and different
values of $q$ ranging from 0.2 to 1.  We obtain very good agreement
between numerical and analytical results for these values of $\te /
\ta$, even for rather large values of $e_{\rm i,eq} \sim 0.5$. 

Equation~(\ref{eq:eiq}) cannot be used for larger values of $\te /
\ta$ as it would then give values of $\ei$ too high for the second order
analysis to be valid.  Anyway, as will be discussed below, larger values 
of $\te/\ta$ are not realistic in the
context of planets migrating in discs.

\subsubsection{Eccentricities of the planets when the system enters an inclination--type resonance}

In the analysis conducted in section~\ref{sec:ana}, we found that
inclination--type resonance would occur for $\ei \equiv e_{\rm i, res}
\sim 0.3$ (eq.~[\ref{eires_small}] and fig.~\ref{fig:eres}).  As we
have already noted, our analysis, which gives the rate of change of
$\varpi_{\rm i,o}$ and $\Omega_{\rm i,o}$ only to zeroth--order in
$e_{\rm i,o}$, is not valid in the regime of large eccentricities.
The numerical simulations presented here do indeed confirm that there
is a low eccentricity regime with $e_{\rm i, res} \simeq 0.3$, which
corresponds to $ \phi_1$ and $\phi_2$  librating about 
  $0^{\circ}$ and $180^{\circ}$, respectively.  In addition, 
  they show that there
is a large eccentricity regime with $e_{\rm i, res} \simeq 0.6$, which
corresponds to $\phi_1$ and  $\phi_2$ both librating about $0^{\circ}$.

The analysis also predicted that inclination--type resonance would
occur for $\eo \equiv e_{\rm o, res} \simeq 0.1$ for $q=0.7$
(eq.~[\ref{eores_small}]).  This value should not be used in the
regime $\phi_2=0$, as in that case $\ei$ is large when the resonance
is triggered and the equations are therefore not developped up to
sufficient order in the eccentricities.  In the low eccentricity
regime, the value found numerically for $e_{\rm o, res}$, 0.18, is
 almost twice as large as the value derived analytically.  In the
large eccentricity regime, the simulations give a similar value of
$e_{\rm o, res} \sim 0.2$.

\subsubsection{Evolution of the inclinations}

After the system enters an inclination--type resonance, according to
equation~(\ref{eq:siso}), the inclinations evolve while maintaining a
constant ratio $\Io / \Ii = q \sa$ (where we have used $s_{\rm i,o} \
\simeq I_{\rm i,o}/2$, valid for small inclinations).

\begin{figure}
\begin{center}
  \includegraphics[scale=0.7,width=0.6\textwidth,height=0.5\textwidth]
  {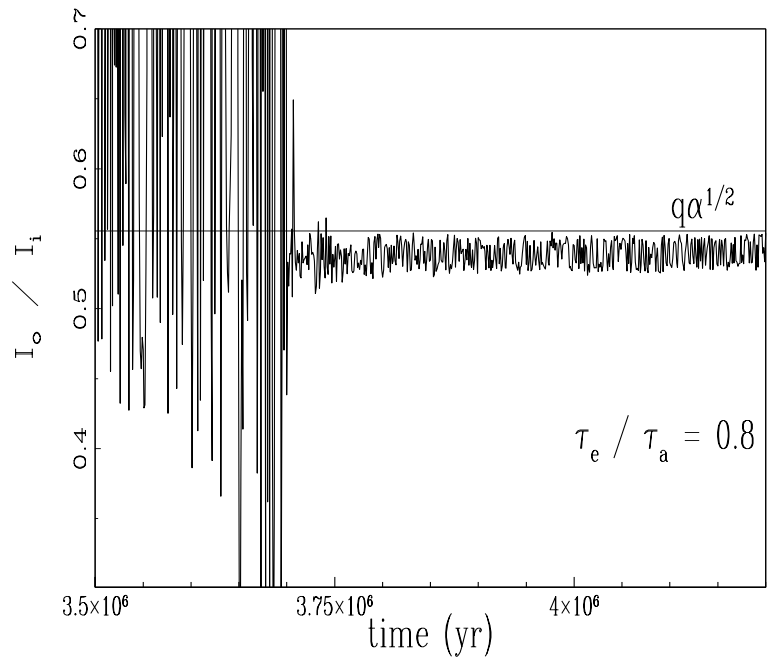}
\end{center}
\caption{$\Io / \Ii$ {\em versus} time (in years) for $q=0.7$ and
  $\te/\ta=0.8$ in the case $\tei=\teo \equiv \te$.  At $t\simeq
  3.7\times 10^6$ yr, the system enters an inclination--type
  resonance.  The horizontal line represents $q \sa$, which is the
  value of $\Io / \Ii$ expected from the analysis.  There is a
  reasonably good agreement between the numerical and analytical
  results.}
\label{fig:siso}
\end{figure}

Numerically, we find this relation to be reasonably well satisfied.
This is illustrated for $\te/\ta=0.8$ and $q=0.7$ in
figure~\ref{fig:siso}, which shows the evolution of $\Io / \Ii$ after
the system has entered an inclination--type resonance.  We have also
verified that the ratio of the inclinations is essentially independent
of $\ta$ and $\te$ for the values of $\te/\ta$ for which the analysis
is valid.

\subsection{Conditions for the onset of inclination--type resonance}
\label{sec:map}

In section~\ref{sec:ana}, we derived a necessary condition
(eq.~[\ref{eq:condres}]) for the system to enter an inclination--type
resonance, which can be written as $\te / \ta > 4 e_{\rm i,res}^2
\left( 1 + q \sa \right)$.  With $e_{\rm i,res}= 0.3$, the right--hand
side of this inequality varies between 0.35 and 0.65 as $q$ varies
between 0 and 1, whereas it varies between 1.4 and 2.6 for $e_{\rm
  i,res}= 0.6$.  We now investigate whether this condition agrees with
numerical results.

\begin{figure}
\begin{center}
\includegraphics[scale=1]{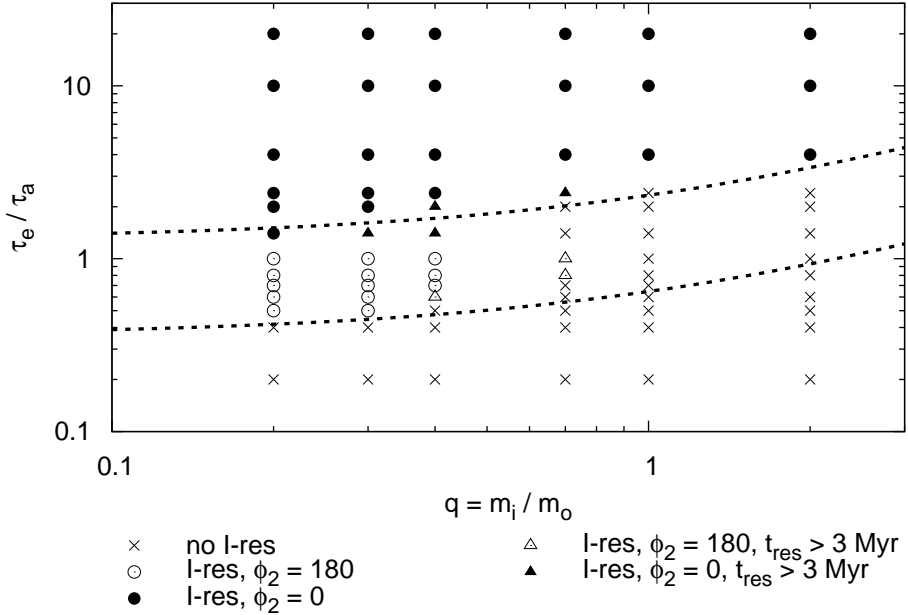}
\end{center}
\caption{Occurrence of inclination--type resonance as a function of
  $q$ and $\te/\ta$ in systems evolved between 0 and 6~Myr.  Here
  $\tei=\teo \equiv \te$. {\em Crosses} represent systems that did not
  enter an inclination--type resonance, {\em open} and {\em filled
    symbols} represent systems that entered an inclination--type
  resonance with $(\phi_1, \phi_2) = (0^{\circ},180^{\circ})$ or
  $(\phi_1, \phi_2) = (0^{\circ},0^{\circ})$, respectively.  {\em
    Circles} and {\em triangles} represent systems that entered an
  inclination--type resonance before or after $t=3$~Myr, respectively.
  The {\em dashed lines} are the curves $\te / \ta = 4 e_{\rm i,res}^2
  \left( 1 + q \sa \right)$ with $e_{\rm i,res}=0.3$ ({\em lower
    line}) and $e_{\rm i,res}=0.57$ ({\em upper line}).  In agreement
  with theoretical expectations, systems in between those two lines
  entered an inclination--type resonance with $(\phi_1, \phi_2) =
  (0^{\circ},180^{\circ})$ if evolved long enough, whereas systems
  above the upper line entered an inclination--type resonance with
  $(\phi_1, \phi_2) = (0^{\circ},0^{\circ})$.}
\label{fig:mapecc}
\end{figure}

We performed a series of runs with $q=$0.2, 0.3, 0.4, 0.7, 1 and 2 and
$\te/ \ta=$0.2, 0.4, 0.5, 0.6, 0.7, 0.8, 1, 1.4, 2, 2.4, 4, 10 and 20.
We integrated the equations for up to $t = 6 \times 10^{6}$ years,
which is longer than or comparable to the age of the disc.  For each
run, we indicate in figure~\ref{fig:mapecc} whether the system entered
an inclination--type resonance or not. Crosses represent runs in which
there was no inclination--type resonance, open and filled symbols
represent runs in which there was an inclination--type resonance with
$(\phi_1, \phi_2) = (0^{\circ},180^{\circ})$ or $(\phi_1, \phi_2) =
(0^{\circ},0^{\circ})$, respectively.  Circles and triangles represent
systems that entered an inclination--type resonance before or after $t=
3 \times 10^6$~years, respectively.

On this diagram, the lower dashed line is the curve $\te / \ta = 4
e_{\rm i,res}^2 \left( 1 + q \sa \right)$ (analytical
condition~[\ref{eq:condres}]) with $e_{\rm i,res}=0.3$.  For $q \le
0.7$, this curve approximates very well the boundary between systems
which do not enter an inclination--type resonance and systems which do
enter such a resonance with $(\phi_1, \phi_2) =
(0^{\circ},180^{\circ})$, i.e. while maintaining small eccentricities.
These results confirm the analytical result $e_{\rm i,res} \simeq 0.3$
(eq.~[\ref{eires_small}] and fig.~\ref{fig:eres}).  We note that, as
$q$ gets larger (heavier inner planet), it takes longer for an
inclination--type resonance to be excited, which is consistent with
the fact that it takes longer for the eccentricity of the inner planet
to grow.  This probably explains why systems above the dashed curve
have not entered an inclination--type resonance for the largest values
of $q$.  We would expect these systems to enter such a resonance if we
carried on the integration beyond 6~Myr.

The upper dashed line is the curve $\te / \ta = 4 e_{\rm i,res}^2
\left( 1 + q \sa \right)$ (analytical condition~[\ref{eq:condres}])
with $e_{\rm i,res}=0.57$, value which gives the best fit to the
boundary between runs which enter an inclination--type resonance for
$(\phi_1, \phi_2) = (0^{\circ},180^{\circ})$ and those which enter the 
resonance for $(\phi_1, \phi_2)
= (0^{\circ},0^{\circ})$.  The fact that this boundary can be fitted
with this curve indicates that the analysis leading to 
the expression~(\ref{eq:eiq}) of the equilibrium value of $\ei$ (which is
used in deriving eq.~[\ref{eq:condres}]) is still valid in the regime
of rather large eccentricities, as already noted in
section~\ref{sec:eieq}, and also that $e_{\rm i,res}$ is essentially
independent of $q$ and $\te/\ta$ in the large eccentricity regime (see
also Thommes \& Lissauer 2003).

\subsection{Influence of varying $t_a$ and $\mo$ }
\label{sec:condchoice}

In the simulations presented above, we fixed $\mo =1$~$\MJ$ and
$t_a=7\times10^{5}$~years (consistent with type~II migration timescale
at a few AU from the star).  Then $q \equiv \mi/\mo$ and $\te/\ta
\equiv t_e / t_a$ were varied. We have checked that the results reported above are unaffected
if we take $t_a$ to be $5\times10^5$ or $10^6$~years and $\mo$ to be
0.5~$\MJ$.

\begin{figure}
\begin{center}
\includegraphics[scale=0.7]{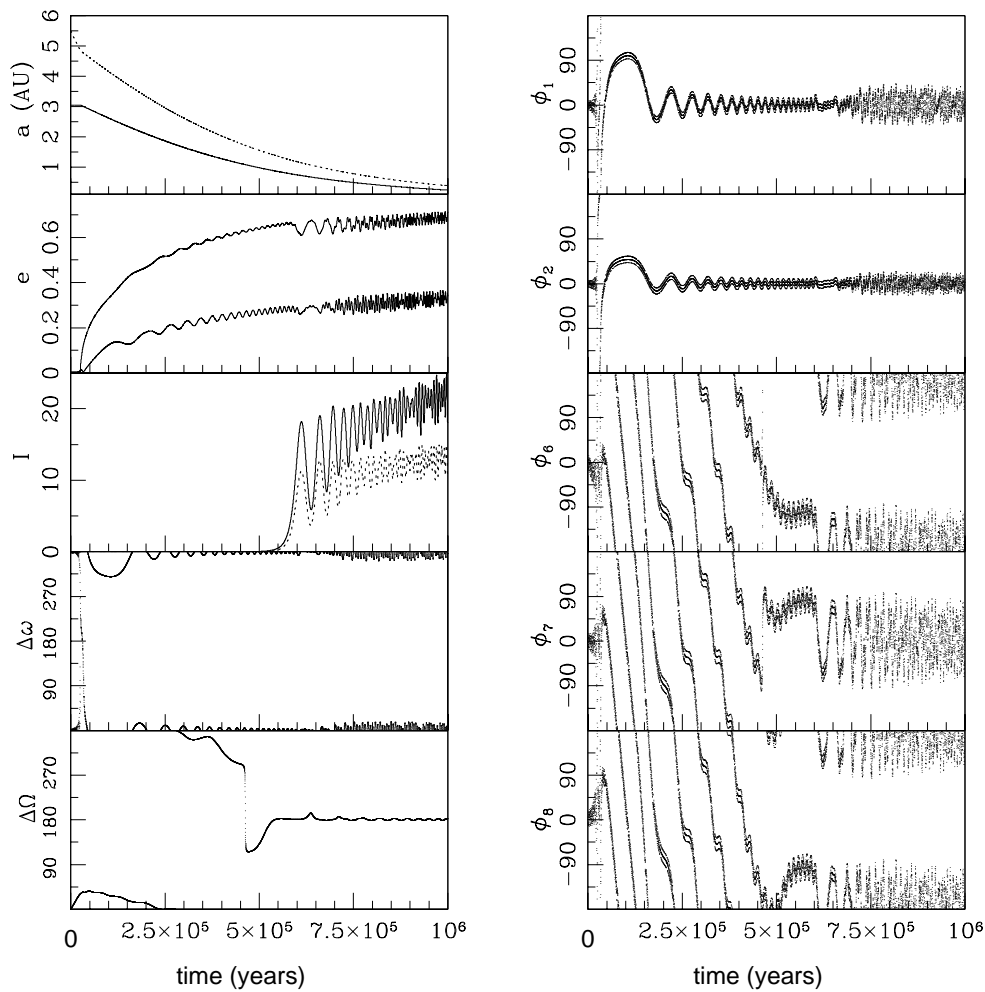}
\end{center}
\caption{Same as figure~\ref{fig:weakte} but for $\mo=\mi=0.05$~$\MJ$
  and $t_a= 2 \times 10^5$~years (and $\te/\ta=4$ as in
  fig~\ref{fig:weakte}).  The evolution is very similar to that observed in
  figure~\ref{fig:weakte} for two Jupiter mass planets.}
\label{fig:nept}
\end{figure}

The \textit{Kepler} mission has found planets with a mass comparable
to or lower than that of Neptune to be very common.
Figure~\ref{fig:nept} shows the evolution of two Neptune mass planets
($\mo=\mi=0.05$~$\MJ$) with $t_a= 2 \times 10^5$~years and
$\te/\ta=4$.  As in the case of two Jupiter mass planets illustrated
above, the system enters an inclination--type resonance when $\ei \sim
0.6$. 

\subsection{Case where the eccentricity of the inner planet is not damped}
\label{sec:cavity}

So far, we have assumed that both planets were interacting with the
disc so that both eccentricities were damped ($t_{ei}$ and $t_{eo}$
both finite).  In this section, we consider the case where the planets
evolve in a cavity, with only the outer planet maintaining contact
with the disc.  Eccentricity damping therefore acts only on the outer
planet ($t_{ei} \to \infty$ and $t_{eo}$ finite).  This situation may
arise if the planets clear--out a gap which is deep enough that the
parts of the disc interior to the orbits are no longer replenished
efficiently from the outer disc.  The inner disc accretes onto the
star over a viscous timescale.  If the outer planet migrates in over a
similar timescale, pushing in the inner planet, contact between the
disc and the inner planet is maintained.  However, is the outer planet
migrates over a slower timescale, which happens when the mass of the
planet is larger than that of the disc in the vicinity of its orbit,
the inner planet loses contact with the disc.  Note that, if the pair
of planets enters an inner cavity carved, e.g., by photoevaporation,
migration of the outer planet ceases and eccentricities cannot grow
anymore, which would prevent the onset of inclination--type resonance.

\begin{figure}
\begin{center}
\includegraphics[scale=0.7]{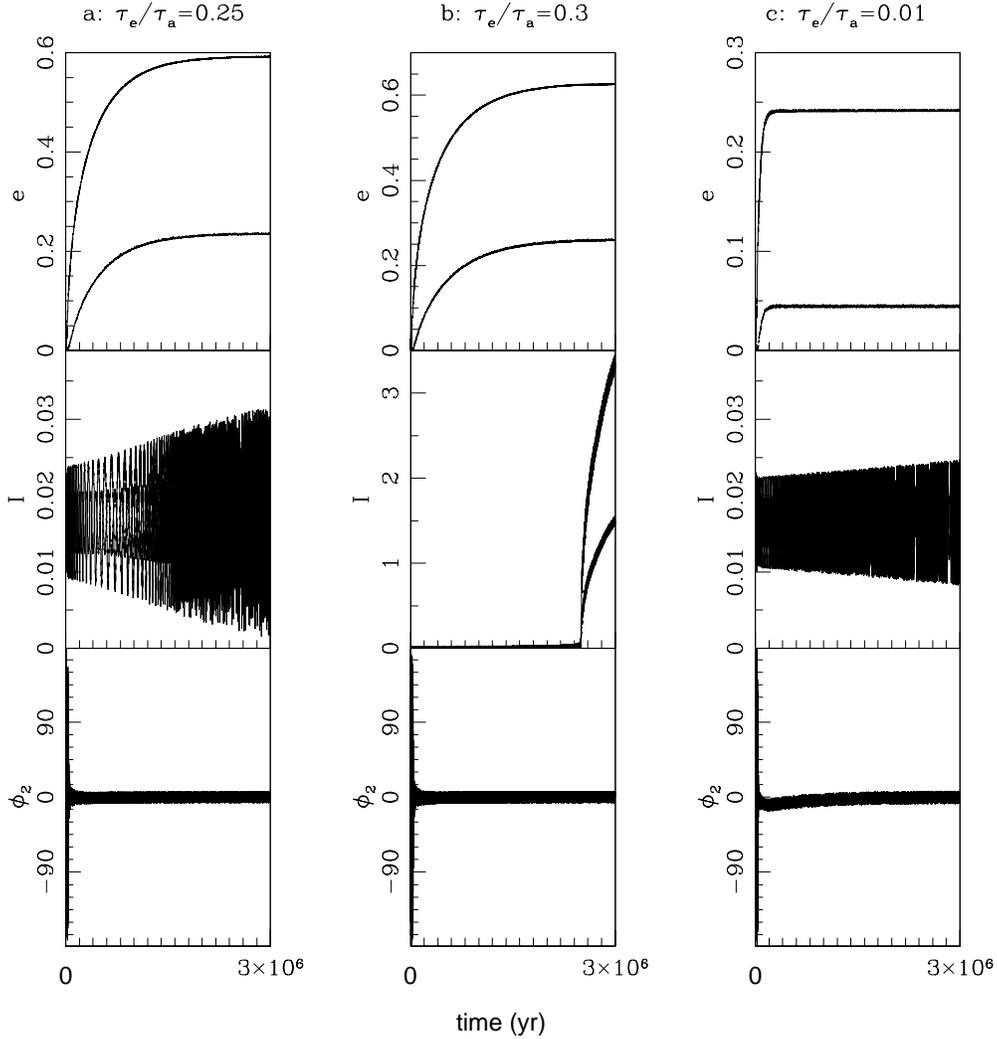}
\end{center}
\caption{Evolution of a system in 2:1 mean motion resonance for
  $q=0.7$ and $\tei \to \infty$. Here $\te \equiv \teo$.  {\it From
    top to bottom:} Eccentricities and inclinations (in degrees) of
  the two planets and resonant angle $\phi_2$ (in degrees) {\it
    versus} time (in years) for $\tau_e/\tau_a=0.25$ ({\it left
    column}), 0.3 ({\it middle column}) and 0.01 ({\it right column}).
  In the plots displaying the eccentricities, the upper and lower
  curves correspond to $\ei$ and $\eo$, respectively.  The case with
  $\tau_e/\tau_a=0.25$ is the same as that displayed on figure
  \ref{fig:strongte} except for $\tei$ which is infinite here. }
\label{fig:notorque}
\end{figure}

We perform the same numerical simulations as above except that now
$\mbox{\boldmath$\Gamma$}_{\rm d,i}$ in equation~(\ref{eq:eomi}) is
set to zero, whereas $\mbox{\boldmath$\Gamma$}_{\rm d,o}$ in
equation~(\ref{eq:eomo}) remains unchanged. Figure~\ref{fig:notorque} shows the eccentricities, inclinations and
resonant angle $\phi_2$ for $\te / \tau_a =0.25, 0.3$ and 0.01, where 
$\te \equiv \teo$.

The case $\te / \tau_a =0.25$ is the same as that shown on
figure~\ref{fig:strongte} with a finite $\tei$.  When $\tei \to
\infty$, the eccentricities reach much larger values.  However, $\ei$
saturates just below 0.6, which is not sufficient to trigger an
inclination--type resonance with
$(\phi_1,\phi_2)=(0^{\circ},0^{\circ})$.  Note that $\ei$ passed
through the value 0.3, which we found to be, in principle, sufficient
to trigger an inclination--type resonance with
$(\phi_1,\phi_2)=(0^{\circ},180^{\circ})$. However, when $\ei=0.3$,
$\eo$ is only about 0.05, below the value $e_{\rm o, res}$ required
for the resonance to be excited.

For $\te / \tau_a =0.3$, the inclination--type resonance is not
excited when the eccentricity of the inner planet is damped (see
figure~\ref{fig:mapecc}).  Here however, we see that $\ei$ grows to
large values and eventually reaches 0.6, at which point an
inclination--type resonance with
$(\phi_1,\phi_2)=(0^{\circ},0^{\circ})$ is triggered.  Therefore, in
this case, suppressing eccentricity damping on the inner planet
significantly affects the evolution of the system.

Figure~\ref{fig:notorque} also shows the case $\te / \tau_a =0.01$.
Damping of the outer planet eccentricity is too strong for either
$e_{\rm i,res}$ or $e_{\rm o,res}$ to be reached. We see on figure~\ref{fig:notorque} that the values at which $\eo$
saturates agree very well with the expectation from
equation~(\ref{eq:eoq}).

The value of $e_{\rm o,res}$ we have calculated analytically is not
valid in the regime $\phi_2=0^{\circ}$ which is observed here.  In
this regime however, numerical simulations performed for finite values
of $\tei$ and $q=0.7$ indicate that $e_{\rm o,res} \simeq 0.2.$ We see
on figure~\ref{fig:eores} that $\eo$ cannot reach this value for $\te
/ \ta < 0.2$, which is in good with the fact that
inclination--type resonance is observed for $\te / \ta = 0.3$ but not
for $\te / \ta = 0.25$ or 0.01.

\subsection{Effect of inclination damping}
\label{sec:idamp}

So far, we have ignored damping of the inclinations due to interaction
with the disc.  For the range of inclinations and eccentricities
considered here, inclination and eccentricity damping are expected to
occur over similar timescales (Bitsch et al. 2013 for gap opening
planets, Cresswell et al. 2007 for non gap opening planets).
Inclination damping can therefore be taken into account by adding the
following term on the right--hand side of equations~(\ref{gammadi})
and~(\ref{gammado}) (Papaloizou \& Larwood 2000):

\begin{equation}
  -\frac{2}{t_i}\left(\dtt{\textbf{r}_{\rm i,o}} \cdot
    \textbf{e}_z\right)\textbf{e}_z
\end{equation}

\noindent where $\textbf{e}_z$ is the unit vector perpendicular to the
disc midplane, and $t_i$ the inclination damping timescale, which is
of the same order as the eccentricity damping timescale.  We have run
a few simulations which show that the addition of this extra force
does not strongly affect the dynamics of the system.
Inclination--type resonances are found to occur for the same values of
the parameters as above, and are characterized by libration of the
resonant angles about the same values as when there is no inclination
damping.

Inclination damping has little effect because it happens over the same
timescale as eccentricity damping.  When $t_e$ is small,
inclination--type resonances do not occur, and therefore inclination
damping is irrelevant.  On the contrary, when $t_e$ is large enough
that inclination--type resonances are excited, inclination damping is
too weak to affect the sudden increase of the inclinations.  It would however
limit the growth of the inclinations  over a timescale longer than the time 
during which the simulations were run.

%
%

\section{Discussion}
\label{sec:discussion}

\subsection{Summary of the main results}

In this paper, we have studied analytically the evolution of the
eccentricities of a pair of planets locked in a 2:1 mean motion
resonance.  In the early stages of the evolution, the planets are in
an eccentricity--type resonance, in which the orbits are in the plane
of the disc.  We have derived the equilibrium eccentricity $e_{\rm i,
  eq}$ reached by the inner planet after a time large compared to the
eccentricity damping timescale (eq.~[\ref{eq:eiq}]) in the case where
the eccentricities of both planets are damped.  In the case where only
the eccentricity of the outer planet is damped, we have calculated its
equilibrium value $e_{\rm o, eq}$ (eq.~[\ref{eq:eoq}]).

We have shown that, for the system to enter an inclination--type
resonance, the eccentricity of the inner planet has to reach a value
$e_{\rm i, res} \sim 0.3$, independent of the migration and
eccentricity damping timescales and only weakly dependent on the mass
ratio $q$ for $q \le 1$ (eq.~[\ref{eires_small}] and
fig.~\ref{fig:eres}). Numerically, we have also shown that there is
another, larger, value of $e_{\rm i, res} \simeq 0.6$ (which was found
by Thommes \& Lissauer 2003).  When the system enters an
inclination--type resonance with $e_{\rm i, res} \simeq 0.3$ ({\em
  small eccentricity} regime), the resonant angles $\phi_1$ and
$\phi_2$ librate about $0^{\circ}$ and $180^{\circ}$, respectively.
In the {\em large eccentricity} regime ($e_{\rm i, res} \simeq 0.6$),
$\phi_1$ and $\phi_2$ both librate about $0^{\circ}$.

We have also derived analytically the value $e_{\rm o, res}$ that the
eccentricity of the outer planet has to reach for an
inclination--type resonance to be excited.  In the low eccentricity
regime, we find $e_{\rm o, res} \simeq 0.1$ for a mass ratio $q
\lesssim 1$.  This value is somewhat smaller than that found
numerically, which is $\simeq 0.2$.  In the large eccentricity regime,
the numerical simulations also give $e_{\rm o, res} \simeq 0.2$ for $q
\lesssim 1$.

If $\ei$ reaches $e_{\rm i, eq}$ while $\eo$ is still below
$e_{\rm o, eq}$, the system keeps evolving in the eccentricity--type
resonance.  As the planets approach each other during their convergent
migration, $\eo$ increases and may at some point reach $e_{\rm o,
  eq}$.  The system may then enter an inclination--type resonance with
either $\ei \simeq 0.3$ or $\simeq 0.6$ if the eccentricity of the
inner planet has continued to increase.

Necessary conditions for the system to enter an inclination--type
resonance are $e_{\rm i, res} < e_{\rm i, eq}$ and $e_{\rm o, res} <
e_{\rm o, eq}$.  This leads to a condition on the ratio of the
eccentricity damping timescale to the semimajor axis damping
timescale, $t_e/t_a$, as a function of the mass ratio $q=\mi/\mo$
(eq.[\ref{eq:condres}] and fig.~\ref{fig:eres} for the case where both
eccentricities are damped and fig.~\ref{fig:eores} for the case where
only the eccentricity of the outer planet is damped).  {\em For $q \le
  1$ and when both eccentricities are damped, we find that the system
  cannot enter an inclination--type resonance if $t_e/t_a <0.2$. This
  result still holds when only the eccentricity of the outer planet is
  damped, at least for $q \lesssim 1$.}

\subsection{Implication for extrasolar planetary systems}

Whether or not the orbit of an extrasolar planet embedded in a disc
and locked in a mean motion resonance with a heavier outer companion
may become inclined due to an inclination--type resonance depends on
whether the eccentricity of the planet can become as high as 0.3.
This in turn depends (weakly) on the mass ratio $q$ and (strongly) on
$t_e/t_a$.

For a wide range of planet masses, the eccentricity damping timescale
due to planet/disc interaction is on the order of a hundred orbits
(see Papaloizou \& Larwood 2000, Cresswell et al. 2007 and Bitsch \&
Kley 2010 for masses $\sim 1-10$~M$_{\oplus}$, 20~M$_{\oplus}$ and
0.1--1~$\MJ$, respectively), i.e. shorter than $10^3$~years for a
planet at a few AU from the central star.  As type~I and type~II
migration timescales at this location are on the order of $10^5$~years,
this implies that $t_e/t_a \sim 10^{-2}$, much smaller than the value
needed for $e_{\rm i, res}$ to be reached. 

Even when the eccentricity of the inner planet is not damped, and for
$q \lesssim 1$, disc torques acting on the outer planet are still too
strong for enabling its eccentricity to reach the value required for
the onset of an inclination--type resonance.

Lee \& Peale (2002) calculated the value of $\tau_e/\tau_a$ consistent
with the observed eccentricities of the two giant planets around
GJ~876, which are found to be in a 2:1 mean motion resonance, assuming
the system had migrated inward.  They found that in the case where the
semimajor axes and eccentricities of both planets were assumed to be
damped, $\tau_e/\tau_a$ had to be about 0.1, whereas it had to be
about 0.01 in the case where only the semimajor axis and eccentricity
of the outer planet were damped.  According to the results presented
in this paper, in neither of these cases could an inclination--type
resonance be triggered.

We conclude that the excitation of inclination through the type of
resonance described here is very unlikely to happen in a system of two
planets migrating in a disc.  If the eccentricities of both planets are
damped, this conclusion does not depend on the mass ratio of the
planets.  If only the eccentricity of the outer planet is damped, this
conclusion holds for at least $q \lesssim 1$.  This is consistent with
the fact that orbits in the multiplanet systems detected by {\em
  Kepler} seem to have both low inclinations (Fabrycky et al. 2012)
and low eccentricities (Kane et al. 2012).  It is also therefore very
unlikely that inclination--type resonance is the cause of the orbital
inclination observed for a number of hot Jupiters.

\section*{Acknowledgements}

We thank our anonymous referee for helpful comments and suggestions
that improved the manuscript.

%
%


%
%

\appendix

\section{Coefficients in the disturbing function}

\label{app:coef}

In table~\ref{tab:coeffs}, we give the expression of the coefficients
$f_i \; (i=1, \ldots, 8)$ and $K_i \; (i=1, \ldots, 4)$ which enter the
expression of the direct part of the disturbing function in
equations~(\ref{RDsec}) and~(\ref{RDres}).  We denote $b_s^{(j)}
(\alpha) $ the Laplace coefficient defined by:

\begin{equation}
  b_{s}^{(j)}(\alpha) = \frac{1}{\pi}\int_0^{2\pi}
  \frac{\cos(j\psi)}{\left(1-2\alpha\cos\psi+\alpha^2\right)^s} \; 
  \text{d}\psi,
\end{equation}

\noindent where $j$ is an integer and $s$ is a half integer.  We
define $\DD \equiv {\rm d}/{\rm d} \alpha$.  In
table~\ref{tab:coeffs}, we also give the numerical value of the $f_i
\; (i=1, \ldots, 8)$ and $K_i \; (i=1, \ldots, 4)$ evaluated at the
2:1 mean motion resonance, where $\alpha=2^{-2/3}$.

\begin{table}
\begin{tabular}{ccc}
  \hline
  \hline
  Coefficient & Expression & Numerical value at \\
  &                    & the 2:1 resonance \\
  \hline
  $f_1$ & $ - \frac{1}{2} \left( 4+ \alpha \DD \right) b_{1/2}^{(2)}$ 
  & -1.19049  \\  \\ [-1.5ex]
  $f_2$ & $ \frac{1}{2} \left( 3 + \alpha \DD \right) b_{1/2}^{(1)}$ 
  & 1.68831     \\ \\ [-1.5ex]
  $f_2^m$ & $ f_2 - 2\alpha$ & 0.42839     \\ \\ [-1.5ex]
  $f_3$ & $ \frac{1}{8} \left( 44 + 14 \alpha \DD + \alpha^2 \DD^2 \right)
  b_{1/2}^{(4)}$ & 1.69573  \\  \\ [-1.5ex]
  $f_4$ & $- \frac{1}{4} \left( 42 + 14 \alpha \DD + \alpha^2 \DD^2 \right)
  b_{1/2}^{(3)}$ & -4.96685 \\  \\ [-1.5ex]
  $f_5$ & $ \frac{1}{8} \left( 38 + 14 \alpha \DD + \alpha^2 \DD^2 \right)
  b_{1/2}^{(2)}$ & 3.59380 \\  \\ [-1.5ex]
  $f_6$ & $ \frac{1}{2} \alpha b_{3/2}^{(3)}$ & 0.81988   \\ \\ [-1.5ex]
  $f_7$ & $ - \alpha b_{3/2}^{(3)}$ & -1.63976     \\ \\ [-1.5ex]
  $f_8$ & $ \frac{1}{2} \alpha b_{3/2}^{(3)}$ & 0.81988     \\ \\ [-1.5ex]
  $K_1$ & $\frac{1}{8} \left( 2 \alpha \DD + \alpha^2 \DD^2 \right)
  b_{1/2}^{(0)}$ & 0.38763 \\  \\ [-1.5ex]
  $K_2$ & $\frac{1}{4} \left( 2 - 2 \alpha \DD - \alpha^2 \DD^2 \right)
  b_{1/2}^{(1)}$ & -0.57570 \\  \\ [-1.5ex]
  $K_3$ & $ - \frac{1}{2} \alpha b_{3/2}^{(1)}$ & -1.55051    \\ \\ [-1.5ex]
  $K_4$ &  $\alpha b_{3/2}^{(1)}$ & 3.10102    \\
  \hline
  \hline
\end{tabular}
\label{tab:coeffs}
\caption{Expression and numerical value at the 2:1 mean motion 
  resonance of the coefficients which enter the expression of the disturbing function}
\end{table}


\section{Damping forces}
\label{sec:forces}

In the $N$--body simulations performed in this paper, the damping
force due to the interaction between the outer planet and the disc is
given by equation~(\ref{gammado}), which we can rewrite under the
form:

\begin{equation}
  \mbox{\boldmath$\Gamma$}_{\rm d}  =  -\left( \frac{1}{t_m}  + \frac{2 }{t_e} \right) \dot{r} \;
  \textbf{e}_r - \frac{1 }{t_m} r  \dot{\theta} \; \textbf{e}_\theta ,
\label{gammado_bis}
\end{equation}

\noindent where we have dropped the subscript 'o' which we have used
for the outer planet. Here $r$ and $\theta$ are the polar coordinates
referred to the central star,  $\textbf{e}_r$ and
$\textbf{e}_\theta$ are the unit vectors along and perpendicular to
the radius vector $\textbf{r}$, respectively, in the orbital plane, and 
the dot denotes a time--derivative.

We now show that adding this force to the equation of motion is
equivalent to adding the terms $- a/t_a - 2 a e^2 / t_e$, with
$t_a=t_m/2$, in the expression of $d a / dt$ and the term $- e /
t_e$ in the expression of $d e / dt$, as done in
section~\ref{sec:migration} (see eq.~[\ref{daodt}] and~[\ref{deodt}]).

The total energy and angular momentum vector per unit mass of the
planet are $E=-G \ms /(2a)$ and $\textbf{H}= r^2 \dot{\theta}
\textbf{e}_z$, respectively, where $\textbf{e}_z =\textbf{e}_r \times
\textbf{e}_\theta$.  Because the planet is acted on by the
perturbative force~(\ref{gammado_bis}), $E$ and $\textbf{H}$ vary with
time according to:

\begin{eqnarray}
\dot{E}  & = &  \dot{\textbf{r}} \cdot \mbox{\boldmath$\Gamma$}_{\rm d}, 
 =  - \left( \frac{1}{t_m}  + \frac{2 }{t_e} \right) \dot{r}^2 
   - \frac{1 }{t_m} (r  \dot{\theta})^2, 
   \label{dEnergy} \\
  \dot{\textbf{H}}  & = & \textbf{r} \times \mbox{\boldmath$\Gamma$}_{\rm d},
  =  - \frac{1 }{t_m} r^2  \dot{\theta} \textbf{e}_z.
  \label{dH}
\end{eqnarray}

The rate of change of $a$ and $e$ is related to that of $E$ and $H$
through (e.g., Burns~1976):

\begin{eqnarray}
\frac{da}{dt} & = & \frac{2 a^2}{G \ms} \dot{E},  \label{da} \\
\frac{de}{dt} & = & \frac{1}{2e} (e^2-1) \left( 2 \frac{\dot{H}}{H}
+ \frac{\dot{E}}{E} \right) . \label{de}
\end{eqnarray}

Substituting equation~(\ref{dEnergy}) into equation~(\ref{da}), we obtain:

\begin{equation}
\left< \frac{da}{dt} \right>   =  \frac{2 a^2}{G \ms} \left[
- \left( \frac{1}{t_m}  + \frac{2 }{t_e} \right) \left< \dot{r}^2 \right>
   - \frac{1 }{t_m} \left< (r  \dot{\theta})^2 \right>
\right], 
\label{da2}
\end{equation}

\noindent where the brackets denote time--averaging over an orbital period.
To perform the time--averaging, we first write $\dot{r}$ and $r \dot{\theta}$ in 
terms of the true anomaly $f$:

\begin{eqnarray}
\dot{r} & = & \frac{na}{\sqrt{1-e^2}} e \sin f , \\
r \dot{\theta} & = & \frac{na}{\sqrt{1-e^2}}  \left(1 + e \cos f \right).
\end{eqnarray}

\noindent We then use the expansion of $\cos f$ and $\sin f$ 
 in terms of 
the mean anomaly $M$, which is 2$\pi$--periodic and a linear function of time.
To obtain $\left< da/dt \right>$ and  $\left< de/dt \right>$ to second--order
in $e$, we only need to expand $\cos f$ and $\sin f$  to first--order in $e$.  This 
gives:

\begin{eqnarray}
\cos f & = & \cos M + e \left( \cos 2 M - 1 \right), \\
\sin f & = & \sin M + e \sin 2 M,
\end{eqnarray}

\noindent so that $\left< \dot{r}^2 \right> = n^2 a^2 e^2 /2$ and 
$\left< (r  \dot{\theta})^2 \right>=n^2 a^2 (1-e^2/2)$.  Substituting 
into equation~(\ref{da2}) and using $G \ms = n^2 a^3$, we 
finally obtain:

\begin{equation}
\left< \dtt{a} \right> = -2a \left( \frac{1}{t_m}+\frac{e^2}{t_e} \right).
\label{da3}
\end{equation}

To calculate the rate of change of $e$, we now substitute 
equations~(\ref{dH}) and~(\ref{da}) into equation~(\ref{de}), and 
use the expression of $E$ and $H$ to obtain:

\begin{equation}
\left< \frac{de}{dt} \right>   =  \frac{1}{2e} (e^2-1) \left( - \frac{2}{t_m} 
-\frac{1}{a} \left< \frac{da}{dt} \right> \right) .
\end{equation}

\noindent Substituting equation~(\ref{da3}), this gives:

\begin{equation}
\left< \frac{de}{dt} \right>   = -\frac{1}{t_e} e.
\label{de2}
\end{equation}

Therefore, as anticipated, adding the damping force~(\ref{gammado_bis})
to the equation of motion is equivalent to adding the terms in 
equations~(\ref{da3})
and~(\ref{de2}) to the rate of change of $a$ and $e$.

\end{document}